\documentclass[a4paper,11pt]{article}
\pdfoutput=1 
 \usepackage{jcappub}
\usepackage{amsmath,amssymb}
\usepackage{bm}
\usepackage{xcolor}
\usepackage{graphicx}
\usepackage{enumitem}
\usepackage{geometry}
\usepackage{xspace}
\usepackage{pdflscape}
\usepackage{placeins}
\usepackage{epstopdf}
\usepackage{comment}
\makeatletter
\gdef\@fpheader{}
\g@addto@macro\bfseries{\boldmath}
\makeatother

\newcommand{\OmegaGW}{\Omega_{\mathrm{GW}}}
\newcommand{\rhoGW}{\rho_{\mathrm{GW}}}

\let\oldsqrt\sqrt
\def\sqrt{\mathpalette\DHLhksqrt}
\def\DHLhksqrt#1#2{%
\setbox0=\hbox{$#1\oldsqrt{#2\,}$}\dimen0=\ht0
\advance\dimen0-0.2\ht0
\setbox2=\hbox{\vrule height\ht0 depth -\dimen0}%
{\box0\lower0.4pt\box2}}

\newcommand{\dd}{\mathrm{d}}
\newcommand{\ee}{e}

\newcommand{\sss}[1]{{\scriptscriptstyle{#1}}}
\newcommand{\boldmathsymbol}[1]{{\ensuremath{\boldsymbol{#1}}}}

\newcommand{\uPl}{\mathrm{Pl}}

\newcommand{\umax}{\mathrm{max}}

\newcommand{\ueff}{\mathrm{eff}}

\newcommand{\usssPl}{\sss{\uPl}}

\newcommand{\ud}{\mathrm{d}}

\newcommand{\calH}{\mathcal{H}}

\newcommand{\Mpc}{\mathrm{Mpc}}

\newcommand{\Mp}{M_\usssPl}

\newcommand{\beq}{\begin{equation}}
\newcommand{\eeq}{\end{equation}}
\newcommand{\bea}{\begin{equation}\begin{aligned}}
\newcommand{\eea}{\end{aligned}\end{equation}}

\newlength{\wsingfig}
\newlength{\wdblefig}
\newlength{\wquadfig}
\newlength{\wtriplefig}
\newcommand{\Eq}[1]{Eq.~(\ref{#1})}

\newcommand{\Fig}[1]{Fig.~{\ref{#1}}}

\newcommand{\Sec}[1]{Sec.~\ref{#1}}

\newcommand{\App}[1]{Appendix~\ref{#1}}

\newcommand{\Hc}[1]{\mathcal{H}}

\renewcommand{\Hc}{\mathcal{H}}
\def\br{\bar\rho}
\def\bp{\bar p}
\def\del{\delta}
\def\ups{\upsilon}
\def\doi{http://doi.org}

\date{today}

\title{Scalar induced gravitational waves from primordial black hole Poisson fluctuations in $f(R)$ gravity}

\author[a]{Theodoros Papanikolaou}

\author[a,b]{Charalampos Tzerefos}

 \author[a,c,f]{Spyros Basilakos}

\author[a,d,e]{Emmanuel N. Saridakis}

\affiliation[a]{National Observatory of Athens, Lofos Nymfon, 11852 Athens, 
Greece}

\affiliation[b]{Department of Physics, National \& Kapodistrian University of Athens, Zografou Campus GR 157 73, Athens, Greece}

\affiliation[c]{ Academy of Athens, Research Center for Astronomy and Applied Mathematics, Soranou Efesiou 4, 11527, Athens, Greece}

\affiliation[d]{CAS Key Laboratory for Researches in Galaxies and Cosmology, 
Department of Astronomy, University of Science and Technology of China, Hefei, 
Anhui 230026, P.R. China}
 
 \affiliation[e]{School of Astronomy, School of Physical Sciences,
University of Science and Technology of China, Hefei 230026, P.R. China}

\affiliation[f]{School of Sciences, European University Cyprus, Diogenes Street, Engomi, 1516 Nicosia, Cyprus}

\emailAdd{papaniko@noa.gr}
\emailAdd{ chtzeref@phys.uoa.gr}
\emailAdd{ svasil@academyofathens.gr}
\emailAdd{msaridak@noa.gr}

\abstract{ 
The gravitational potential of a gas of initially randomly distributed primordial black holes (PBH) can induce a stochastic gravitational-wave (GW) background through second-order gravitational effects. This GW background can be abundantly generated in a cosmic era dominated by ultralight primordial black holes, with masses $m_\mathrm{PBH}<10^{9}\mathrm{g}$. In this work, we consider $f(R)$ gravity as the underlying gravitational theory and we study its effect at the level of the gravitational potential of Poisson distributed primordial black holes. After a general analysis, we focus on the $R^2$ gravity model. In particular, by requiring that the scalar induced GWs (SIGWs) are not overproduced, we find an upper bound on the abundance of PBHs at formation time $\Omega_\mathrm{PBH,f}$ as a function of their mass, namely that $\Omega_\mathrm{PBH,f}<5.5\times 10^{-5}\left(\frac{10^9\mathrm{g}}{m_\mathrm{PBH}}\right)^{1/4}$, which is $45\%$ tighter than the respective upper bound in general relativity. Afterwards, by considering $R^2$ gravity as an illustrative case study of an $f(R)$ gravity model, we also set upper bound constraints on its mass parameter $M$. These mass parameter constraints, however, should not be regarded as physical given the fact that the Cosmic Microwave Background (CMB) constraints on $R^2$ gravity are quite tight. Finally, we conclude that the portal of SIGWs associated to PBH Poisson fluctuations can act as a novel complementary probe to constrain alternative gravity theories. 
}


\begin{document}
\maketitle

\section{Introduction}

Primordial black holes (PBHs), firstly proposed in the early `70s~\cite{1967SvA....10..602Z, Carr:1974nx,1975ApJ...201....1C,1979A&A....80..104N}, are formed in the early universe before the birth 
of stars, out of the collapse of overdensity regions whose energy density perturbations are higher than a critical threshold~\cite{Harada:2013epa,Musco:2018rwt,Papanikolaou:2022cvo}. They are
currently attracting an increasing attention since they can address a number of issues of modern cosmology. According to recent arguments, they can 
potentially account for a part or all of the dark matter content of the Universe
\cite{Chapline:1975ojl}, and additionally they can offer an explanation for     
the large-scale structure formation through Poisson 
fluctuations \cite{Meszaros:1975ef,Afshordi:2003zb}. Furthermore, they can 
provide seeds for the supermassive black holes residing in the centre of 
galaxies ~\cite{1984MNRAS.206..315C, Bean:2002kx}, as well as constitute viable 
candidates for the progenitors of the black-hole merging events recently 
detected by the LIGO/VIRGO collaboration~\cite{LIGOScientific:2018mvr} through 
the emission of gravitational waves (GWs). Other evidence in favor of the PBH 
scenario can be found  in \cite{Clesse:2017bsw}.

Due to the significance  of PBHs and the  huge progress achieved in the field 
of gravitational-wave astronomy, there have been many attempts connecting 
PBHs and GWs~\cite{Sasaki:2018dmp}. Firstly, a large amount of research has 
been devoted to the GW background signals associated to PBH merging events 
 \cite{Nakamura:1997sm, Ioka:1998nz, 
Eroshenko:2016hmn, Raidal:2017mfl, Zagorac:2019ekv,Hooper:2020evu}. Moreover, 
extensive research has been also performed regarding the PBH Hawking radiated-graviton background~\cite{Anantua:2008am,Dong:2015yjs} as well as concerning the scalar induced GWs (SIGWs)
connected to the primordial high curvature perturbations which gave rise to 
PBHs ~\cite{Saito_2009,Bugaev:2009zh, Nakama_2015, Yuan:2019udt, Braglia:2020eai,Zhou:2020kkf,Fumagalli:2020nvq} (for a recent 
review see \cite{Domenech:2021ztg}).
However, apart from the aforementioned GW 
signals, it has been recently noted in~\cite{Papanikolaou:2020qtd}, and 
further studied in~\cite{Domenech:2020ssp,Kozaczuk:2021wcl}, that the Poisson 
fluctuations of a gas of randomly distributed PBHs can induce second-order 
GWs at distances much larger than the PBH mean separation scale.  These GWs are 
not induced by the primordial curvature perturbations, which gave rise to 
PBHs, but instead by the PBH density fluctuations themselves and can be abundantly produced during an early PBH dominated era naturally driven by ultralight PBHs, which evaporate before BBN time ~\cite{GarciaBellido:1996qt, Hidalgo:2011fj, Martin:2019nuw, Zagorac:2019ekv}.

At the same time, there are many reasons indicating that one should construct 
modified gravitational theories. At the theoretical level, gravitational 
modifications are known to be able to improve the renormalizability issues of 
general relativity \cite{Stelle:1976gc,Addazi:2021xuf}. At the phenomenological level, modified 
gravity can offer an alternative way to explain the two phases of the Universe's 
accelerated expansion, namely the early-time, inflationary one 
\cite{Nojiri:2010wj,Martin:2013tda}, 
and/or the late-time, dark-energy one 
\cite{CANTATA:2021ktz,Capozziello:2011et,Cai:2015emx}. In all cases, these 
modified gravitational theories possess general relativity as a 
particular limit, but in general they have a richer structure and extra 
degrees of freedom that can describe the Universe's evolution. 

One of the simplest classes of modified gravity is $f(R)$ gravity, which is 
obtained through the extension of the Einstein-Hilbert Lagrangian to an 
arbitrary function of the Ricci scalar \cite{DeFelice:2010aj}. Apart from its 
general cosmological application, in the inflationary framework the 
particular subclass of the theory known as Starobinsky, or $R^2$ gravity 
\cite{Starobinsky:1980te}, proves to be one of the best-fitted models to the cosmological
data~\cite{Planck:2018jri}. Hence, due to its success, $f(R)$ gravity has 
been extensively studied in the literature. In particular, in such 
investigations one is in general interested in extracting the corrections on 
various observational signals, induced by the $f(R)$ modifications  on 
top of the corresponding general-relativity predictions  (see 
\cite{Nojiri:2006gh,Amendola:2006we,Bean:2006up,Faulkner:2006ub,Hu:2007nk,
Starobinsky:2007hu, 
Cognola:2007zu,Brax:2008hh,Appleby:2009uf,Olmo:2011uz,Cai:2013lqa,Leon:2014yua,Ohta:2015efa,
Capozziello:2015yza,Nunes:2016drj,Astashenok:2017dpo,Capozziello:2018ddp,
Naik:2018mtx, 
Elizalde:2018rmz,Arnold:2019zup,Gogoi:2020ypn,Tang:2020sjs,Wilson:2020uzk,
Toniato:2021vmt} and references 
therein). 

Therefore, in the present work we are interested in investigating the GW signal 
induced by PBH Poisson fluctuations, in the framework of $f(R)$ gravity. In 
particular, since all the relevant studies up to now have been performed in the 
framework of general relativity, apart from~\cite{Chen:2021nio,Kawai:2021edk,Lin:2021vwc} where the authors study the primordial SIGWs in modified gravity constructions, in the following we calculate   the effect of $f(R)$ 
corrections on the PBH gravitational potential power spectrum and subsequently on the associated SIGW background. In this way, one 
may use it as an extra and novel method to constrain on the one hand the PBH abundances and on the other hand possible $f(R)$ modifications, constituting in this way an independent test of general relativity.

The plan of the work is as 
follows: In \Sec{GRPBH}, we review the PBH gravitational potential in 
general relativity and in \Sec{fRPBH} we perform the extended 
analysis, extracting the PBH gravitational potential in the framework of $f(R)$ 
gravity. In \Sec{SIGW}, we make a case study within $f(R)$ gravity theories and extract the relevant SIGW signal focusing on the simplest $f(R)$ gravity model, namely the $R^2$ gravity, treating in this way its mass parameter $M$ as a free parameter. Then, in \Sec{ConstrSrarob} by demanding that SIGWs are not overproduced at PBH evaporation time, we obtain, on the one hand, upper bound constraints on the PBH abundance at formation time $\Omega_\mathrm{PBH,f}$ as a function of the PBH mass $m_\mathrm{PBH}$ and, on the other hand, upper bounds on the mass parameter $M$ of $R^2$ gravity as a function of $m_\mathrm{PBH}$ and $\Omega_\mathrm{PBH,f}$. Finally, 
\Sec{Conclusions} is devoted to the conclusions.
 
\section{The primordial black hole gravitational potential in general relativity}\label{GRPBH}

In the context of general relativity (GR),  the action is written as follows:
\begin{equation} \label{GRaction}
S = \frac{1}{16\pi G} \int d^4x \sqrt{-g} \,(R-2\Lambda)  + \int d^4x \sqrt{-g} 
\mathcal{L}_{\mathrm{m}} ,
\end{equation}
with $G$ being the gravitational Newton constant (throughout this paper we work in units where $c=1$), $R$ the Ricci scalar, $\Lambda$ the cosmological constant,  $\mathcal{L}_{\mathrm{m}}$ 
  the total matter Lagrangian density (radiation, baryonic and dark matter) of the 
Universe and 
$\, T^{\mathrm{m}}_{\mu \nu} \equiv -\frac{2}{\sqrt{-g}}\frac{\delta 
\mathcal{L}_{\mathrm{m}}}{\delta g^{\mu \nu}}$ the corresponding total matter energy-momentum tensor. 
Varying the action (\ref{GRaction}) with respect to the metric  $g^{\mu \nu}$ 
we obtain the usual Einstein field equations, namely
\begin{equation}
 R_{\mu \nu} - \frac{1}{2} g_{\mu \nu} R + \Lambda g_{\mu \nu}  = 8\pi G T^{\mathrm{m}}_{\mu \nu}. \label{gre}
 \end{equation}
 Note that the Bianchi 
identity $\nabla_{\mu} G^{\mu}_{\nu}= 0 $ implies the conservation of the total 
energy-momentum tensor. 

\subsection{Background evolution}

Proceeding to a cosmological setup, we consider a flat 
 Friedmann - Lema\^itre - Robertson -Walker (FLRW) background metric of the form
\begin{equation}
\label{FLRWmetric_background}
ds^{2}_\mathrm{b}=-dt^{2}+a^{2}(t)\delta_{ij}dx^{i}dx^{j}\,,
\end{equation}
where $a(t)$ is the scale factor.  By adopting this background metric and assuming that the total matter content of the Universe is described by the perfect fluid energy-momentum tensor $T^{\mathrm{m}}_{\mu\nu} = \mathrm{diag}( - \br , \bp , \bp , \bp)$, where $\br$ and $\bp$ are the total matter (i.e. including radiation, baryonic and dark matter) energy density and pressure respectively,
the GR field equations give rise to the two Friedmann equations:
    \begin{align}
    H^2&= \frac {8\pi G  }{3} \br + \frac{\Lambda}{3} \equiv \frac {8\pi G  }{3} \br_{\mathrm{tot}} \label{F1}\\
\dot{H} + H^2 &=-\frac{4\pi G  }{3}  \left(\br + 3\bp \right) + \frac{\Lambda}{3} \equiv -\frac{4\pi G  }{3}  \left(\br_{\mathrm{tot}} + 3\bp_{\mathrm{tot}}\right), \label{F2}
    \end{align}
where $H=\dot{a}/a$ is the Hubble parameter, with dots denoting derivatives with respect to the cosmic time $t$. In the above expressions  $\br_{\mathrm{tot}}$ and $\bp_{\mathrm{tot}}$ correspond to the total background energy density and pressure of the Universe, i.e the total matter sector as well as   the cosmological constant term, which is interpreted as a dark energy fluid with $\rho_{\mathrm{de}}= - p_{\mathrm{de}} \equiv \frac{\Lambda}{8\pi G}$ whose energy-momentum tensor is $ T^{\mathrm{de}}_{\mu\nu} = \mathrm{diag}( - \rho_{\mathrm{de}}, p_{\mathrm{de}}, p_{\mathrm{de}}, p_{\mathrm{de}})$. Nevertheless, since in this work we focus on the early-time matter (i.e. PBH) dominated era, the contribution of the cosmological constant or effective dark energy at the background level can be neglected.

Lastly, it proves convenient to introduce the conformal time  $\eta$ defined through $\mathrm{d}t \equiv a \mathrm{d}\eta$, and similarly the conformal Hubble parameter defined as $\Hc \equiv a'/a=aH$,
where primes denote derivatives with respect to  $\eta$. Hence, the above two Friedmann equations become simply
    \begin{align}
     \Hc^{2} &=\frac {8\pi G a^2}{3} \br_\mathrm{tot} \label{Fr1000}\\
\Hc ' &=-\frac{4\pi G a^2}{3}  (\br_\mathrm{tot} + 3\bp_\mathrm{tot}). \label{FR2000}
    \end{align}

 \subsection{Scalar perturbations}
 \label{GRperts}

Let us now refer to the perturbation evolution.  Focusing on scalar 
perturbations, the perturbed FLRW metric  in the Newtonian gauge reads as 
\bea
\label{perturbed FLRW metric with scalar perturbations}
\mathrm{d}s^2 = a^2(\eta)\left\lbrace-(1+2\Psi)\mathrm{d}\eta^2  + \left[(1-2\Phi)\delta_{ij}\right]\mathrm{d}x^i\mathrm{d}x^j\right\rbrace \, ,
\eea
where for convenience we perform the calculations using the conformal time $\eta$.
In the above ansatz,
$\Psi$ and $\Phi$  stand for the Bardeen potentials  \cite{Bardeen:1980kt},
which are first order quantities in cosmological perturbation theory. 

Further, we allow perturbations around the background stress-energy tensor of the total matter content of the Universe (matter and radiation) which we write as follows:  
\begin{align}
 T^{0}_{0} &= -(\br + \del \rho)\nonumber \\
 T^{0}_{i} &= (\br + \bp)\ups_{i} \, , \, \, \ups_{i} \equiv a \delta u_{i} \nonumber\\
 T^{i}_{j} &= \bp ( \del ^{i}_{j} + \Pi^{i}_{j}), \label{hydrodynamicstressenergytensor}
\end{align}
where $\del \equiv \del \rho / \br $ \, is the relative energy density perturbation, $\delta u_i \equiv \ups_i/a $ is the velocity perturbation and $\Pi^i_j$ is the (dimensionless) anisotropic stress. The evolution of $\Phi$ and  $\Psi$ is governed by the perturbed Einstein equations, which are \cite{LL}: 
   \begin{align}
      3\Hc(\Phi' + \Hc\Psi) -\nabla ^2\Phi &= -4\pi G a^2 \, \delta \rho \label{E1} \\
        (\Phi' + \Hc\Psi)_{,i} &= 4\pi G a^2 (\br + \bp) \ups_{,i} \label{E2} \\ 
         \Phi'' + \Hc(\Phi' + 2\Psi') + (\Hc^2 + 2\Hc')\Phi + \nabla ^2(\Phi - \Psi) /3   &= 4\pi G a^2  \delta p \label{E3} \\
    \Phi - \Psi  &= 8\pi G a^2 \bp \Pi . \label{E4}
    \end{align}
    
During the time period we are concerned with, namely before BBN, the anisotropic stress of the Universe is negligible since we do not have the presence of free-streaming particles. Thus, from (\ref{E4}) we see that $\Phi \approx \Psi$, which we will adopt from now on. This potential can actually be identified with the PBH gravitational potential, whose behavior will be derived in the following analysis.  \footnote{The first-order gravitational potential due to the primordial energy density perturbations is ignored here as we concentrate on the induced GW signal due to the PBH energy density perturbations.  This contribution can be added to the contribution calculated in our work, if we desire to include  the primordial SIGWs  \cite{Domenech:2021ztg} too. } 

We proceed by defining the total entropy perturbation as 
\beq
\mathcal{S} \equiv \Hc \left( \frac{\delta p}{\bp '} - \frac{\delta \rho}{\br '} \right) \label{S}.
\eeq
Since the (total) energy-momentum tensor is conserved, the background continuity equation holds, namely $\bar{\rho} ' = - 3\Hc (\bar{\rho} + \bar{p})$. Therefore, from (\ref{S}) we acquire:
\beq
\delta p = c^2_s [\delta \rho - 3 (\br + \bp) \mathcal{S}], \label{delP}
\eeq
where $w \equiv \bp / \br $ is the equation-of-state parameter and $ c^2_\mathrm{s}\equiv \bp' / \br' $ is the sound speed square of the total matter content of the Universe. Finally, one can combine (\ref{E1}) with (\ref{E3}) and   (\ref{delP}) to get the following  equation governing the behavior of the gravitational potential $\Phi$ :
\beq\label{eq:Phi:GR}
\Phi'' + 3\mathcal{H}\left(1+c^2_\mathrm{s}\right)\Phi' - c^2_\mathrm{s}\nabla^2\Phi + 3\left(c^2_\mathrm{s} - w \right)\mathcal{H}^2\Phi = - \frac{9}{2}c^2_\mathrm{s} (1+ w) \mathcal{H}^2\mathcal{S}.
\eeq
\subsection{The Power Spectrum of the PBH Gravitational Potential}\label{sec:PowerSpectrumPhiGR}

Having extracted  above the background and the pertrubation equations for the PBH gravitational potential, we derive here the corresponding power spectrum following closely \cite{Papanikolaou:2020qtd}. As it is standardly adopted in the literature, we assume that PBHs are formed in the radiation-dominated (RD) era.  Hence,  considering PBHs as a matter fluid,  their formation process can be regarded as a transition of a fraction of the radiation energy density into PBHs.  Thus,  assuming that PBHs are randomly distributed in space at formation time,  their energy density is inhomogeneous while the total energy density of the background is homogeneous.  Consequently,  the PBH energy density perturbation can be viewed as an isocurvature Poisson fluctuation. 
As it was found in ~\cite{Papanikolaou:2020qtd}, the Poissonian power spectrum 
for the PBH density contrast, assuming monochromatic PBH mass function~\cite{MoradinezhadDizgah:2019wjf}, reads as
\beq
\label{eq:PowerSpectrum:PBH}
\mathcal{P}_\delta(k) = \frac{k^3}{2\pi^2}P_\delta(k)= \frac{2}{3\pi} 
\left(\frac{k}{k_{\mathrm{UV}}}\right)^3 \Theta(k_\mathrm{UV}-k),
\eeq
where $k_{\mathrm{UV}}\equiv a/\bar{r}$ is a UV cut-off scale related to the 
mean PBH separation scale.  This UV cut-off scale is introduced here since at scales smaller than the mean PBH separation scale   the PBH fluid description is not valid. In particular,  at these scales one probes the granularity of the PBH energy density field entering the non-linear regime where $\mathcal{P}_\delta(k)>1$. Straightforwardly, one can show that the UV cut-off scale reads as~\cite{Papanikolaou:2020qtd}
\beq
k_\mathrm{UV} = \mathcal{H}_\mathrm{f}\Omega^{1/3}_\mathrm{PBH,f},
\eeq
where $\mathcal{H}_\mathrm{f}$ and $\Omega_\mathrm{PBH,f}$ are respectively the conformal Hubble parameter and the PBH abundance at PBH formation time.

Then the next step is to relate the above power spectrum of the PBH energy density perturbations to the power spectrum for the PBH gravitational potential $\Phi$.  In order to achieve this we should have in mind that since in the RD era,  $\Omega_\mathrm{PBH}\equiv 
\frac{\rho_\mathrm{PBH}}{\rho_\mathrm{tot}}\propto a$, hence if the initial abundance of PBHs is large enough, then PBHs can potentially dominate the 
Universe energy budget. Consequently, the isocurvature PBH  energy density perturbation in the RD era will be converted to an adiabatic curvature perturbation in the subsequent PBH dominated era~\cite{Kodama:1986fg,Kodama:1986ud}, which will be related to a gravitational potential $\Phi$. 

To derive now $\Phi$ from $\delta_\mathrm{PBH}$,  we use as an intermediate variable  the uniform-energy density curvature perturbation of a fluid, $\zeta$, which is related 
with the Bardeen potential $\Phi$ and the respective 
energy density perturbation by the following definition \cite{Wands:2000dp}:
\begin{equation}
    \zeta \equiv -\Phi - \Hc \frac{\delta \rho}{\bar{\rho}'}.
\end{equation}
If the total energy-momentum tensor is conserved, the (background) continuity equation  $\bar{\rho} ' = - 3\Hc (\bar{\rho} + \bar{p})$ holds, and thus $\zeta $ is expressed as
\begin{equation}
  \zeta \equiv -\Phi + \frac{\delta}{3(1+w)}, \label{z}
\end{equation}
where $ w \equiv \bar{p}/\bar{\rho}$ is the equation-of-state parameter of the total matter content of the Universe. In our case, since the energy-momentum tensors of radiation and PBH-matter are separately conserved,  we can use (\ref{z}) for
$\zeta_\mathrm{r}$ and $\zeta_\mathrm{PBH}$ and acquire:
\beq\label{eq:zeta_r}
\zeta_\mathrm{r}=-\Phi+\frac{1}{4} \delta_\mathrm{r},
\eeq
\beq\label{eq:zeta_PBH}
\zeta_\mathrm{PBH}=-\Phi+\frac{1}{3} \delta_\mathrm{PBH}.
\eeq
Finally, we introduce the isocurvature perturbation defined as:
\beq\label{eq:S definition}
S = 3\left(\zeta_\mathrm{PBH}-\zeta_\mathrm{r}\right) =\delta_\mathrm{PBH} - 
\frac{3}{4} \delta_\mathrm{r} \,.
\eeq

On superhorizon scales, $\zeta_\mathrm{r}$ and  $\zeta_\mathrm{PBH}$ are 
conserved separately~\cite{Wands:2000dp},  like the isocurvature perturbation 
$S$. Thus, in the PBH-dominated era, $\zeta\simeq \zeta_\mathrm{PBH} = 
\zeta_{\mathrm{r}}+S/3 \simeq S/3$. Since $S$ is conserved, it can be calculated 
at formation time $t_\mathrm{f}$.  Therefore, neglecting  the adiabatic 
contribution associated to the radiation fluid at the PBH formation time, since it is negligible for the scales considered here,  from 
\Eq{eq:S definition} we obtain that $S=\delta_\mathrm{PBH}(t_\mathrm{f})$. Hence, we finally find
\bea
\label{eq:zeta:delta:superH}
\zeta\simeq \frac{1}{3} \delta_\mathrm{PBH}(t_\mathrm{f})\quad \mathrm{if}\quad 
k\ll \mathcal{H}\,. 
\eea
Using now the fact that $\zeta\simeq -\mathcal{R}$ on superhorizon scales (see 
e.g. \cite{Wands:2000dp}), where $\mathcal{R}$ is the comoving curvature 
perturbation defined by
\bea
\label{eq:zeta:Bardeen}
\mathcal{R}  = 
\frac{2}{3}\frac{{\Phi}^\prime/\mathcal{H}+\Phi}{1+w }+\Phi\, ,
\eea
one gets straightforwardly that in the PBH-matter dominated era, where $w=0$ and 
$\Phi$ is constant in time \cite{Wands:2000dp}, 
\bea
\label{eq:Phi:delta:superH00}
\Phi\simeq -\frac{1}{5} \delta_\mathrm{PBH}(t_\mathrm{f})\quad \mathrm{if}\quad 
k\ll \mathcal{H}\,.
\eea

On sub-Hubble scales, one can determine the evolution of $\delta_
\mathrm{PBH}$ by solving the evolution equation for the matter density 
perturbations,  namely the M\'eszaros growth equation~\cite{Meszaros:1974tb},  
which,  in the case of a Universe with radiation and PBH-matter, takes the   form:
\bea\label{eq:Meszaros in GR}
\frac{\dd^2 \delta_\mathrm{PBH}}{\dd s^2}+\frac{2+3s}{2s(s+1)}\frac{\dd 
\delta_\mathrm{PBH}}{\dd s}-\frac{3}{2s (s+1)} \delta_\mathrm{PBH}=0\,.
\eea
By solving the above   equation one can find that the
the dominant solution deep in the PBH-dominated era can be written as 
$\delta_\mathrm{PBH}\simeq 3s\,  \delta_\mathrm{PBH}(t_\mathrm{f})/2$. Now, the 
relation   between the Bardeen potential and the density contrast is dictated 
by the Poisson equation, and in a matter-dominated era takes the form
\bea
\label{eq:delta:Phi:MD}
\delta_\mathrm{PBH} =  -\frac{2}{3} \left(\frac{k}{\mathcal{H}}\right)^2\Phi.
\eea
Therefore, plugging  the solution for $\delta_\mathrm{PBH}$ into the aforementioned formula, one obtains
\bea
\label{eq:Phi:delta:subH}
\Phi\simeq -\frac{9}{4}\left(\frac{\calH_{\mathrm{d}}}{k}\right)^2\, 
\delta_\mathrm{PBH}(t_\mathrm{f})
\quad \mathrm{if}\quad k\gg {\calH}_{\mathrm{d}}\, ,
\eea
where ${\calH}_{\mathrm{d}}$ is the conformal Hubble function at 
PBH domination time.
Finally,  making an interpolation between (\ref{eq:Phi:delta:subH}) and 
(\ref{eq:Phi:delta:superH00}),  and using (\ref{eq:PowerSpectrum:PBH}) one obtains
that 
\beq\label{eq:PowerSpectrum:Phi:PBHdom}
\mathcal{P}_\Phi(k) \equiv\frac{k^3}{2\pi^2}P_\Phi(k)= 
\frac{2}{3\pi}\left(\frac{k}{k_\mathrm{UV}}\right)^3 
\left(5+\frac{4}{9}\frac{k^2}{k_{\mathrm{d}}^2}\right)^{-2}\, ,
\eeq
where $k_\mathrm{d}\equiv \cal{H}_\mathrm{d}$ is the comoving scale exiting the Hubble radius at PBH 
domination time.  From \Eq{eq:PowerSpectrum:Phi:PBHdom}, one can see that 
$\mathcal{P}_\Phi$ has a broken power-law behavior: when $k\ll k_{\mathrm{d}}$ we have that
$\mathcal{P}_\Phi \propto k^3$, while when $k\gg 
k_\mathrm{d}$ we acquire $\mathcal{P}_\Phi \propto 1/k$.  We mention that it reaches its maximum when $k \sim k_\mathrm{d}$, where 
$\mathcal{P}_\Phi$ is of order $(k_\mathrm{d}/k_{\mathrm{UV}})^3$.

\section{The primordial black hole gravitational potential in $f(R)$ gravity}
\label{fRPBH}

In the previous section we presented the calculation of the  PBH gravitational potential power spectrum in the framework 
of general relativity. In this section we proceed to the bulk of our analysis, which is to
perform the same calculation but in the case of $f(R)$ modified gravity, extracting the corresponding 
corrections. 

We consider   a modified action of the form  \cite{DeFelice:2010aj}
\begin{equation} 
S = \frac{1}{16\pi G} \int d^4x \sqrt{-g} \, f(R) + \int d^4x \sqrt{-g} 
\mathcal{L}_\mathrm{m} \label{fr},
\end{equation}
where $f(R)$ is a general function of the Ricci scalar  $R$. Variation 
of the action (\ref{fr}) with respect to the metric $g^{\mu \nu}$ 
yields the following field equations:
\begin{equation}
 FR_{\mu \nu} - \frac{1}{2} g_{\mu \nu} f + (g_{\mu \nu} \Box - 
\nabla_{\mu}\nabla_{\nu})F = 8\pi G T^\mathrm{m}_{\mu \nu} \label{sf},  
\end{equation}
where we have set $F \equiv \mathrm{d}f(R)/\mathrm{d}R$. One characteristic feature of the richer structure of $f(R)$ gravity is the existence of an additional propagating degree of freedom, the so-called scalaron field \cite{Starobinsky:1980te}. Its equation can be obtained by taking the trace of (\ref{sf}), which yields:
\begin{equation}
    \Box F(R)= \frac{1}{3}\left[ 2f(R) - F(R)R + 8\pi G \, T^\mathrm{m} \right] \equiv \frac{dV}{dF}, \label{scalaron}
\end{equation}
where $T^\mathrm{m}$ is the trace of the energy-momentum tensor of the (total) matter content of the Universe. As we observe,  equation (\ref{scalaron}) is a wave equation for $\phi_{\mathrm{sc}} \equiv F(R)$ whose mass is given by $m_{\mathrm{sc}}^2 \equiv d^2 V/ dF^2$, which reads:
\begin{equation}
    m_{\mathrm{sc}}^2 = \frac{1}{3}\left( \frac{F}{F_{,R}} - R \right), \label{scalmass}
\end{equation}
where $F_{,R} \equiv dF/dR = d^2f/dR^2$. An alternative way to see this is by performing a conformal transformation to the Einstein frame \cite{DeFelice:2010aj}. Amongst others, the presence of this additional degree of freedom induces an extra polarization mode for the gravitational waves  \cite{DeFelice:2010aj}, as we will see in the next section.

For our purposes, we shall formulate $f(R)$ gravity in terms of an effective curvature-induced fluid. Specifically, we shall express the equations (\ref{sf}) as the corresponding ones in GR (\ref{gre}), with the addition of the following 
energy-momentum tensor \cite{ArNes:2019} instead of the one induced by the cosmological constant:  
\begin{align}
   T^\mathrm{f(R) \, \mu }_{\nu} &\equiv (1 - F) R^{\mu}_{\nu} + 
\frac{1}{2}\delta^{\mu}_{\nu}( f - R) - (\delta^{\mu}_{\nu} \Box - 
\nabla^{\mu}\nabla_{\nu})F \label{tfr}.
\end{align} 

Similarly to the GR case, we will first examine the evolution at the background and perturbation levels, and then we will calculate the power spectrum of the PBH gravitational potential.

\subsection{Background evolution}

Applying $f(R)$ gravity to a cosmological framework, namely using     the FLRW metric (\ref{FLRWmetric_background}), we 
extract  the   Friedmann equations, which in terms of the conformal time are written as
    \begin{align}
     \Hc^{2} &=\frac {8\pi G a^2}{3} \br_\mathrm{tot} \label{FR1}\\
\Hc ' &=-\frac{4\pi G a^2}{3}  (\br_\mathrm{tot} + 3\bp_\mathrm{tot}). \label{FR2}
    \end{align}
    We mention that these equations acquire the same form as  in the  GR case, with 
 the only difference being that in the total content of the Universe we need to take into account the contribution of the $f(R)$ curvature-induced effective fluid, whose energy density and pressure are given by \cite{DeFelice:2010aj}:
\begin{align}
 \br_\mathrm{f(R)} &\equiv -T^{\mathrm{f(R)} \,0 }_{0} = \frac{1}{8\pi G a^2} \Big( 3\Hc^2- \frac{1}{2} a ^2 f + 3F\Hc'- 3\Hc F' \Big) \label{reff}\\
 \bp_\mathrm{f(R)} &\equiv \frac{T^{ \mathrm{f(R)} \,i }_{ i}}{3} = \frac{1}{8\pi G a^2} \Big(- 2\Hc'- \Hc^2 + \frac{1}{2} a ^2 f - F\Hc'- 2F \Hc^2 + F'' + \Hc F' \Big). \label{peff}
\end{align}

\subsection{Scalar perturbations}

In order to describe the evolution of scalar perturbations, we shall use again the metric (\ref{perturbed FLRW metric with scalar perturbations}) and the perturbed form of the (total) energy-momentum tensor (\ref{hydrodynamicstressenergytensor}).
The perturbed field equations are similar in form with the corresponding ones of GR, with the addition of $\delta \rho_\mathrm{f(R)}, \delta p _\mathrm{f(R)}, v_\mathrm{f(R)}$ and  $\Pi _\mathrm{f(R)}$. They are provided explicitly in Appendix \ref{AppendixA}. Therefore, we need to take into account the contribution of the $f(R)$ curvature-induced effective fluid to the expressions introduced in subsection \ref{GRperts}.

Within this context, we define the total entropy perturbation as: 
\beq
\mathcal{S}_\mathrm{tot} \equiv \Hc \left( \frac{\delta p_\mathrm{tot}}{\bp_\mathrm{tot} '} - \frac{\delta \rho_\mathrm{tot}}{\br_\mathrm{tot} '} \right). \label{frS}
\eeq
Again the total energy-momentum tensor is conserved, so the background continuity equation holds, namely $\bar{\rho}_\mathrm{tot} ' = - 3\Hc (\bar{\rho}_\mathrm{tot} + \bar{p}_\mathrm{tot})$, so from (\ref{frS}) we acquire:
\beq
\delta p_\mathrm{tot} = c^2_\mathrm{tot} [\delta \rho_\mathrm{tot} - 3 (\br_\mathrm{tot} + \bp_\mathrm{tot}) \mathcal{S}_\mathrm{tot}], \label{dpfr}
\eeq
where $w_\mathrm{tot} \equiv \bp_\mathrm{tot} / \br_\mathrm{tot} $ is the total equation-of-state parameter and $ c^2_\mathrm{tot}\equiv \bp_\mathrm{tot}' / \br_\mathrm{tot}' $ is the sound speed square of the total content of the Universe. By combining (\ref{P1}) with (\ref{P3}) and  (\ref{dpfr}) we get the following equation governing the behavior of the gravitational potential $\Phi$:
\beq\label{eq:Phi:fr}
\Phi'' + 3\mathcal{H}\left(1+c^2_\mathrm{tot}\right)\Phi' - c^2_\mathrm{tot}\nabla^2\Phi + 3\left(c^2_\mathrm{tot} - w_\mathrm{tot} \right)\mathcal{H}^2\Phi = - \frac{9}{2}c^2_\mathrm{tot} (1+ w_\mathrm{tot}) \mathcal{H}^2\mathcal{S}_\mathrm{tot}.
\eeq
 
\subsection{The Power Spectrum of the PBH Gravitational Potential in $f(R)$ gravity}

We can now repeat the procedure  of subsection \ref{sec:PowerSpectrumPhiGR} and extract the power spectrum of the PBH gravitational potential $\mathcal{P}_\Phi$ within the context of $f(R)$ gravity. At this point, we should emphasise that we are agnostic about the production mechanism of PBHs within $f(R)$ gravity. We merely assume that they are formed during RD era after the end of inflation \footnote{Ultralight PBHs may arise as well from the growth of metric perturbations during the matter dominated stage after the end of inflation and before the scalaron decay~\cite{Motohashi:2012tt,Mathew:2020nqo}. However, we do not consider such scenarios in the present work. We focus on PBHs during an RD era as it is standardly assumed in the literature.} and that they are Poisson distributed at formation time, an assumption which is rather reasonable. For this reason, the methodology adopted in this section for the derivation of $\mathcal{P}_\Phi$ is model independent. Then, the setup described in \Sec{SIGW} and \Sec{ConstrSrarob} for the calculation of the SIGW signal and the derivation of constraints on the parameters of our $f(R)$ model at hand, namely the $R^2$ gravity, can be easily generalised to alternative gravitational theory [See e.g in~\cite{Papanikolaou:2022hkg} the generalisation to teleparallel gravity]. 

In the following, we will make use of cosmological perturbation theory working with perturbations in the Jordan frame in order to extract the power spectrum of the PBH gravitational potential\footnote{As it was found in~\cite{Tsamparlis:2013aza,Postma:2014vaa}, it should be emphasized that physics is frame independent and the Jordan and Einstein frame are equivalent giving the same physical observables. One can always define perturbations in both the Einstein and Jordan frames. See e.g.~\cite{DeFelice:2010aj} regarding the definition of the curvature perturbation within $f(R)$ theories of gravity.}.

To begin with, we need to take into account the presence of the $f(R)$ curvature-induced effective fluid. 
Therefore, on top of the usual $\zeta_\mathrm{r}$ and $\zeta_\mathrm{PBH}$, we have $\zeta_\mathrm{f(R)}$, too. Since by construction its energy-momentum tensor (\ref{tfr}) is conserved, we can use (\ref{z}) to get:
\beq
\zeta_\mathrm{f(R)}=-\Phi+\frac{1}{3(1+w_\mathrm{f(R)})} \delta_\mathrm{f(R)} ,
\eeq 
\noindent
where $w_\mathrm{f(R)} \equiv  \bar{p}_\mathrm{f(R)} /\bar{\rho}_\mathrm{f(R)} = \frac{-a^2 f + 2\big((1+2F) \mathcal{H}^{2}  - \mathcal{H} F'+ (2+ F) F' + \mathcal{H}' - F'' \big)}{a^2 f - 6(\mathcal{H}^{2} - \mathcal{H} F' + \mathcal{H}' F )}$ is the equation-of-state parameter  of the effective fluid. Thus, we can study now how these curvature perturbations evolve on super-Hubble ($k\ll \mathcal{H}$) and sub-Hubble ($k\gg \mathcal{H}$) scales.

On super-Hubble scales, $\zeta_\mathrm{r}$ and $\zeta_\mathrm{PBH}$ are separately conserved \cite{Wands:2000dp},  as is the isocurvature perturbation between them, which is defined by
\beq
S=3\left(\zeta_\mathrm{PBH}-\zeta_\mathrm{r}\right)\,.
\eeq
However, the total curvature perturbation is not conserved and is equal to   
\beq
\zeta=-\Phi+\frac {\delta_{\mathrm{tot}}}{3(1+w_\mathrm{tot})} = \frac{\frac{4}{3}\bar{\rho}_\mathrm{r}\zeta_\mathrm{r} + \bar{\rho}_\mathrm{PBH}\zeta_\mathrm{PBH} + (1 + w_\mathrm{f(R)})\bar{\rho}_\mathrm{f(R)} \zeta_\mathrm{f(R)}}{\frac{4}{3}\bar{\rho}_\mathrm{r} + \bar{\rho}_\mathrm{PBH} + (1 + w_\mathrm{f(R)})\bar{\rho}_\mathrm{f(R)}} \label{zeta}.
\eeq
At this point, we should stress that given the fact that we consider that PBHs are formed during the RD era after the end of inflation, it is reasonable to assume that at the background level, the energy contribution from the $f(R)$ fluid will be negligible compared to the contribution of radiation and matter in form of PBHs, i.e. $\bar{\rho}_\mathrm{r}/\bar{\rho}_\mathrm{f(R)} \gg 1$ and $\bar{\rho}_\mathrm{PBH}/\bar{\rho}_\mathrm{f(R)} \gg 1$. In addition, on the scales we are interested in, namely the PBH scales, the dominant contribution to the curvature perturbation during the PBH dominated era, will be due to the PBH curvature perturbation. Thus one can safely neglect $(1 + w_\mathrm{f(R)})\bar{\rho}_\mathrm{f(R)} \zeta_\mathrm{f(R)}$ and $(1 + w_\mathrm{f(R)})\bar{\rho}_\mathrm{f(R)}$  from the numerator and the denominator of \Eq{zeta} respectively. Consequently, $\zeta$ can be recast in the following form: 
 \begin{equation}
     \zeta = \frac{4}{4+3s}\zeta_\mathrm{r}+\frac{3s}{4+3s}\zeta_\mathrm{PBH} ,     
 \end{equation}
 with $ s\equiv \frac{a}{a_\mathrm{d}}$, 
 where $a_\mathrm{d}$ denotes the value of the scale factor $a$ at the time PBHs start to dominate. From this expression, we can see that $\zeta$ evolves from its initial value $\zeta_\mathrm{r}$, deep in the radiation era, to $\zeta_\mathrm{PBH}$, deep in the PBH era.
As a consequence, in the PBH-dominated era, $\zeta\simeq \zeta_\mathrm{PBH} = \zeta_{\mathrm{r}}+S/3$. Since $S$ is conserved  on super-Hubble scales, it can be evaluated at formation time $t_\mathrm{f}$. Furthermore, the isocurvature perturbation can be identified with $\delta_\mathrm{PBH}(t_\mathrm{f})$, which  will be calculated in the  following  subsection, assuming implicitly a uniform radiation energy density in the background. Indeed, in the following  we will focus on the PBH contribution and we will ignore the usual adiabatic contribution (associated to the radiation fluid), which is negligible at the scales we are interested in, hence we simply have
\bea
\label{eq:zeta:delta:superH}
\zeta\simeq \frac{1}{3} \delta_\mathrm{PBH}(t_\mathrm{f})\quad \mathrm{if}\quad k\ll \mathcal{H}\,. 
\eea

Concerning the super-Hubble scales, as we show in Appendix \ref{AppendixB}, in $f(R)$ gravity one can also use the  property   $\zeta\simeq -\mathcal{R}$  (see e.g. \cite{Wands:2000dp}), where $\mathcal{R}$ is the comoving curvature perturbation defined in (\ref{eq:zeta:Bardeen}), by requiring that $\delta F \approx 0 $ \footnote{The assumption $\delta F \approx 0$ is a reasonable one since, as it was checked numerically, $(\Phi-\Psi)/\Phi = \delta F/ (F\Phi)$ is very small during the time period considered here and can vary between $10^{-30}$ up to $10^{-9}$ depending on the choice of the PBH mass $m_\mathrm{PBH}$, the initial PBH abundance $\Omega_\mathrm{PBH,f}$ and the parameter of the underlying gravity theory. See Appendix \ref{app:anisotropic_stress} for more details. } for $k\ll \mathcal{H} $ which ensures (in addition to the usual assumption that the anisotropic stress of the total matter content is negligible at these scales) that $\Psi \approx \Phi$. During a matter-dominated era, such as the one driven by PBHs, $\Phi^\prime $ can be neglected since it is proportional to the decaying mode, thus we obtain  $\mathcal{R}=-\zeta=(5/3)\Phi$. Finally, combining with (\ref{eq:zeta:delta:superH}), this implies that
\bea
\label{eq:Phi:delta:superH}
\Phi\simeq -\frac{1}{5} \delta_\mathrm{PBH}(t_\mathrm{f})\quad \mathrm{if}\quad k\ll \mathcal{H}\,.
\eea

Let us now focus on sub-Hubble scales. One can determine the evolution of $\delta_
\mathrm{PBH}$ by solving the evolution equation of the matter density perturbations,  namely the Meszaros equation~\cite{Meszaros:1974tb},  in a Universe where we have radiation,  matter in form of PBHs, and an effective dark energy   fluid due to the $f(R)$ gravity modulations, which should however be negligible before Big Bang Nucleosynthesis (BBN) time where one expects a subdominant energetic contribution from the dark energy  sector.

At the background level, the Friedman equation (\ref{FR1}) can be expressed as $
\calH^2 = \frac{8\pi Ga^2}{3}\left[\bar{\rho}_\mathrm{PBH}+\bar{\rho}_\mathrm{r}+ \br^\mathrm{f(R)}\right]$
where $\br^\mathrm{f(R)}$ is given by (\ref{reff}), 
where every sector obeys the conservation equation separately  \cite{ArNes:2019}.  Neglecting  then the effective fluid contribution, as justified above,  the Friedmann equation can be recast as 
\beq\label{eq:Friedmann:Meszaros:DE_neglected}
\calH^2 \simeq H^2_\mathrm{f}\Omega^2_\mathrm{PBH,f}\left(\frac{1}{s}+\frac{1}{s^2}\right),
\eeq
where $s\equiv a/a_\mathrm{d}$ and $a_\mathrm{d}$ denotes the time at the transition from the radiation to the PBH domination era, and where we have assumed that $\Omega_\mathrm{r,f}\simeq 1$ since PBHs are considered to be formed in the radiation era \cite{Papanikolaou:2020qtd}. Note that the scale factor is normalised at one at formation time, i.e. $a_\mathrm{f}=1$.  

At the perturbation level, we can use the standard cosmological perturbation theory at subhorizon scales, where the matter perturbations obey the growth equation~\cite{Ma:1995ey,Saridakis:2021qxb}:
\beq\label{eq:growth:equation:GR}
\delta^{\prime\prime}_\mathrm{m}+\calH\delta^\prime_\mathrm{m}-4\pi G a^2\bar{\rho}_\mathrm{m}\delta_\mathrm{m} = 0.
\eeq
Treating the gas of PBHs as a matter fluid and accounting for the screening of the gravitational constant due to $f(R)$ gravity modification, one should replace in the above equation $\delta_\mathrm{m}$ with $\delta_\mathrm{PBH}$ and $G$ with $G_\mathrm{eff}$ defined as~\cite{PhysRevD.76.104043} 
\beq\label{eq:G_eff:f(R)}
G_\mathrm{eff} \equiv \frac{G}{F}\left( \frac{1 +4\frac{k^2}{a^2}\frac{F_\mathrm{,R}}{F}}{1+3\frac{k^2}{a^2}\frac{F_\mathrm{,R}}{F}}\right).
\eeq
Hence, assembling everything, and  using $s$ as  the time variable, the growth equation \eqref{eq:growth:equation:GR} can be recast in the following form:
\bea\label{eq:Meszaros in f(R):subhorizon}
\frac{\dd^2 \delta_\mathrm{PBH}}{\dd s^2}+\frac{2+3s}{2s(s+1)}\frac{\dd \delta_\mathrm{PBH}}{\dd s}-\frac{3}{2s (s+1)} \frac{1}{F} \frac{1 +4\frac{k^2}{a^2}\frac{F_\mathrm{,R}}{F}}{1+3\frac{k^2}{a^2}\frac{F_\mathrm{,R}}{F}}\delta_\mathrm{PBH}=0\,.
\eea
 
We proceed by relating  our solution for $\delta_\mathrm{PBH}$ from (\ref{eq:Meszaros in f(R):subhorizon}) with $\Phi$, via the sub-Hubble scale approximation of the time-time field equation in $f(R)$ gravity for the PBH dominated era (equations (\ref{P1}) and (\ref{dreff}) of Appendix \ref{AppendixA}). At the end, one gets the modified Poisson equation which reads as follows:
\bea\label{eq:Phi:delta:subH:f(R)}
\delta_\mathrm{PBH} = -\frac{2}{3} \left( \frac{k}{\mathcal{H}}\right)^2 \frac{F \left( 1+ 3 \frac{k^2}{a^2} \frac{F_{,R}}{F}\right)}{1+ 2 \frac{k^2}{a^2} \frac{F_{,R}}{F}}\Phi.
\eea
Hence, making   an interpolation between \Eq{eq:Phi:delta:superH} and  \Eq{eq:Phi:delta:subH:f(R)} as in the case of GR, and using the expression for the PBH matter power spectrum in \Eq{eq:PowerSpectrum:PBH} we straighforwardly extract the following PBH gravitational potential power spectrum:
\beq\label{eq:PowerSpectrum:Phi:PBHdom:f(R)}
\mathcal{P}_\Phi(k) \equiv\frac{k^3}{2\pi^2}P_\Phi(k) = \frac{2}{3\pi}\left(\frac{k}{k_\mathrm{UV}}\right)^3\left[5+\frac{2}{3}\left(\frac{k}
{\mathcal{H}}\right)^2\frac{F}{\xi(a)} \left(\frac{1+ 3 \frac{k^2}{a^2} \frac{F_{,R}}{F}}{1+ 2 \frac{k^2}{a^2} \frac{F_{,R}}{F}}\right)\right]^{-2}.
\eeq
In the above expression, $\xi(a)$ is defined as
\beq\label{eq:xi_definition}
\xi(a)\equiv \frac{\delta_\mathrm{PBH}(a)}{\delta_\mathrm{PBH}(a_\mathrm{f})},
\eeq 
where $\delta_\mathrm{PBH}(a)$ is the solution of
\Eq{eq:Meszaros in f(R):subhorizon}. As checked numerically, $\xi(a)$ has a mild 
dependence on on the comoving scale $k$, and thus for practical reasons we will consider $\xi(a)$ as $k$ independent. Lastly, note that in the case of GR we have $F = 1$ and 
$\xi(a)\simeq \frac{3}{2}\frac{a}{a_\mathrm{d}}$, and thus recovering the result of
(\ref{eq:PowerSpectrum:Phi:PBHdom}).


\section{Scalar induced gravitational waves in Starobinsky $R^2$ gravity}
\label{SIGW}
In the previous section we derived the power spectrum of the gravitational potential of initially Poisson-distributed PBHs, therefore in this section we are able to extract the stochastic gravitational wave background induced at second order from the PBH Poisson fluctuations. Since we will perform specific calculations, we have to specify our $f(R)$ form. As we mentioned in the Introduction, one of the most studied cases, which can also give rise to an inflationary scenario with a very efficient agreement with observations, is the Starobinsky or $R^2$ gravity \cite{Starobinsky:1980te}, in which 
\bea\label{StarfR}
 f(R)=R+\frac{R^2}{6M^2},
\eea
with $M$ being the model parameter with dimensions of mass. This mass parameter is well fixed by the amplitude of the curvature power spectrum on CMB scales and it is equal to $M=10^{-5}\Mp$~\cite{Planck:2018jri}. However, given the simplicity of the Starobinsky gravity model - it constitutes the simplest realisation beyond GR within $f(R)$ gravity - we will use it in the following as our case study $f(R)$ gravity model in order to see how one can constrain an $f(R)$ gravity theory using the portal of the SIGWs associated to PBH Poisson fluctuations. Consequently, in the following sections the mass parameter $M$ of Starobinsky gravity will be considered as a free parameter of the underlying gravity theory.

Before deriving the GW spectrum induced from a gas of PBHs,  it is important to highlight here a major issue emerging from the study of induced GWs at second order.  In particular, while the tensor modes are gauge invariant at first order, this is not valid at second order \cite{Hwang_2017,Tomikawa:2019tvi,DeLuca:2019ufz,Yuan:2019fwv,Inomata:2019yww}. This implies that, a priori, one needs to specify in which gauge the gravitational waves are observed. However, in this work we explore a GW backreaction problem without paying attention to observational predictions. In particular, if the energy density associated to the induced gravitational waves overcomes the one of the background, one expects perturbation theory to break down in any gauge. Hence, it is legitimate to assume that our findings bear little dependence on the gauge choice. 

\subsection{Tensor Perturbations}\label{subsec:tensor_perturbations}

Having clarified the gauge choice issue, we continue by studying the tensor perturbations $h_{ij}$ induced by the gravitational potential $\Phi$. In particular, the perturbed metric in the Newtonian gauge, assuming as usual zero anisotropic stress and $\delta F/ F \approx 0$ [See Appendix \ref{app:anisotropic_stress}], is written as
\bea
\label{metric decomposition with tensor perturbations}
\mathrm{d}s^2 = a^2(\eta)\left\lbrace-(1+2\Phi)\mathrm{d}\eta^2  + \left[(1-2\Phi)\delta_{ij} + \frac{h_{ij}}{2}\right]\mathrm{d}x^i\mathrm{d}x^j\right\rbrace,
\eea
where we have multiplied by a factor $1/2$ the second order tensor perturbation as is standard in the literature \footnote{The contribution from the first-order tensor perturbations is not considered here since we concentrate on gravitational waves induced by scalar perturbations at second order.}. Then, by Fourier transforming the tensor perturbations and taking into account the three polarization modes of the GWs in $f(R)$ gravity, namely the $\times$ and the $+$ as in GR and the scalaron one, denoted with $\mathrm{sc}$, the equation of motion for the tensor modes $h_\boldmathsymbol{k}$ reads as
\beq
\label{Tensor Eq. of Motion}
h_\boldmathsymbol{k}^{s,\prime\prime} + 2\mathcal{H}h_\boldmathsymbol{k}^{s,\prime} + (k^{2}-\lambda m^2_\mathrm{sc}) h^s_\boldmathsymbol{k} = 4 S^s_\boldmathsymbol{k}\, ,
\eeq
where $\lambda=0$ when $s = (+), (\times)$ and $\lambda=1$ when $s=(\mathrm{sc})$. The scalaron mass term, $m^2_\mathrm{sc}$, is given by equation (\ref{scalmass}), and thus
in the case of the Starobinsky model it becomes simply $m^2_\mathrm{sc}=M^2$. 
The source function $S^s_\boldmathsymbol{k}$ is given by
\beq
\label{eq:Source:def}
S^s_\boldmathsymbol{k}  = \int\frac{\mathrm{d}^3 \boldmathsymbol{q}}{(2\pi)^{3/2}}e^s_{ij}(\boldmathsymbol{k})q_iq_j\left[2\Phi_\boldmathsymbol{q}\Phi_\boldmathsymbol{k-q} + \frac{4}{3(1+w_\mathrm{tot})}(\mathcal{H}^{-1}\Phi_\boldmathsymbol{q} ^{\prime}+\Phi_\boldmathsymbol{q})(\mathcal{H}^{-1}\Phi_\boldmathsymbol{k-q} ^{\prime}+\Phi_\boldmathsymbol{k-q}) \right],
\eeq
where $s = (+), (\times), (\mathrm{sc})$. The polarization tensors  $e^{s}_{ij}(k)$ are defined as \cite{Capozziello:2011et}
\beq
e^{(+)}_{ij}(\boldmathsymbol{k}) = \frac{1}{\sqrt{2}}
\begin{pmatrix}
1 & 0 & 0\\
0 & -1 & 0 \\ 
0 & 0 & 0 
\end{pmatrix}, \quad
e^{(\times)}_{ij}(\boldmathsymbol{k}) = \frac{1}{\sqrt{2}}
\begin{pmatrix}
0 & 1 & 0\\
1 & 0 & 0 \\ 
0 & 0 & 0 
\end{pmatrix}, \quad 
e^{(\mathrm{sc})}_{ij}(\boldmathsymbol{k}) = \frac{1}{\sqrt{2}}
\begin{pmatrix}
0 & 0 & 0\\
0 & 0 & 0 \\ 
0 & 0 & 1 
\end{pmatrix}.
\eeq


Regarding the time evolution of the potential $\Phi$ considering $c^2_\mathrm{tot} \approx w_\mathrm{tot}$ and neglecting entropic perturbations \Eq{eq:Phi:fr}, can be recast as 
\bea
\label{Bardeen potential 2}
\Phi_\boldmathsymbol{k}^{\prime\prime} + \frac{6(1+w_\mathrm{tot})}{1+3w_\mathrm{tot}}\frac{1}{\eta}\Phi_\boldmathsymbol{k}^{\prime} + w_\mathrm{tot}k^2\Phi_\boldmathsymbol{k} =0\, .
\eea
The above equation accepts a solution with one constant and one decaying mode on super sound-horizon scales.  In the late-time limit, one can neglect the decaying mode,  and write the solution for the Fourier transform of $\Phi$ as $\Phi_\boldmathsymbol{k}(\eta) = T_\Phi(\eta) \phi_\boldmathsymbol{k}$, where $\phi_\boldmathsymbol{k}$ is the value of the gravitational potential at some initial time (which here we consider it to be the time at which PBHs dominate the energy content of the Universe,  $x_\ud$) and $T_\Phi(\eta)$ is a transfer function, defined as the ratio of the dominant mode between the times $x$ and $x_\ud$. 
Consequently, \Eq{eq:Source:def} can be written in a more compact form as
\beq
\label{Source}
S^s_\boldmathsymbol{k}  =
\int\frac{\mathrm{d}^3 q}{(2\pi)^{3/2}}e^{s}(\boldmathsymbol{k},\boldmathsymbol{q})F(\boldmathsymbol{q},\boldmathsymbol{k-q},\eta)\phi_\boldmathsymbol{q}\phi_\boldmathsymbol{k-q}\, ,
\eeq
where
\bea
\label{F}
\!\!\!\!\!
F(\boldmathsymbol{q},\boldmathsymbol{k-q},\eta) & \equiv 2T_\Phi(q\eta)T_\Phi\left(|\boldmathsymbol{k}-\boldmathsymbol{q}|\eta\right)  + \frac{4}{3(1+w)}\left[\mathcal{H}^{-1}qT_\Phi^{\prime}(q\eta)+T_\Phi(q\eta)\right]
\\  & \kern-2em
\ \ \ \ \ \ \ \ \ \ \  \ \ \ \ \ \  \ \ \ \ \ \ \ \ \ \ \ \ \ \ \ \ \ \ \ \ 
\cdot \left[\mathcal{H}^{-1}\vert\boldmathsymbol{k}-\boldmathsymbol{q}\vert T_\Phi^{\prime}\left(|\boldmathsymbol{k}-\boldmathsymbol{q}|\eta\right)+T_\Phi\left(|\boldmathsymbol{k}-\boldmathsymbol{q}|\eta\right)\right],
\eea
and the contraction  $e^s_{ij}(\boldmathsymbol{k})q_iq_j \equiv e^s(\boldmathsymbol{k},\boldmathsymbol{q})$ can be expressed in terms of the spherical coordinates $(q,\theta,\varphi)$ of the vector $\bm{q}$ as 
\beq
e^s(\boldmathsymbol{k},\boldmathsymbol{q})=
\begin{cases}
\frac{1}{\sqrt{2}}q^2\sin^2\theta\cos 2\varphi \mathrm{\;for\;} s= (+)\\
\frac{1}{\sqrt{2}}q^2\sin^2\theta\sin 2\varphi  \mathrm{\;for\;} s= (\times) \\
\frac{1}{\sqrt{2}}q^2\cos^2\theta \mathrm{\;for\;} s= (\mathrm{sc})
\end{cases}
\, .
\eeq
Finally,  the solution of  \Eq{Tensor Eq. of Motion} for the tensor modes $h^s_\boldmathsymbol{k}$ can be obtained using the Green's function formalism where one can write for $h^s_\boldmathsymbol{k}$ that
\bea
\label{tensor mode function}
a(\eta)h^s_\boldmathsymbol{k} (\eta)  =4 \int^{\eta}_{\eta_\mathrm{d}}\mathrm{d}\bar{\eta}\,  G^s_\boldmathsymbol{k}(\eta,\bar{\eta})a(\bar{\eta})S^s_\boldmathsymbol{k}(\bar{\eta}),
\eea
where the Green's function  $G^s_{\bm{k}}(\eta,\bar{\eta})$ is the solution of the homogeneous equation 
\beq
\label{Green function equation}
G_\boldmathsymbol{k}^{s,\prime\prime}(\eta,\bar{\eta})  + \left( k^{2} - \lambda m^2_\mathrm{sc} -\frac{a^{\prime\prime}}{a}\right)G^s_\boldmathsymbol{k}(\eta,\bar{\eta}) = \delta\left(\eta-\bar{\eta}\right),
\eeq
with the boundary conditions $\lim_{\eta\to \bar{\eta}}G^s_\boldmathsymbol{k}(\eta,\bar{\eta}) = 0$ and $ \lim_{\eta\to \bar{\eta}}G^{s,\prime}_\boldmathsymbol{k}(\eta,\bar{\eta})=1$.  

Having extracted above the tensor perturbations, the next step is to derive the tensor power spectrum, $\mathcal{P}_{h}(\eta,k)$ for the different polarization modes, which is defined as the equal time correlator of the tensor perturbations through the following relation:
\bea\label{tesnor power spectrum definition}
\langle h^r_{\boldmathsymbol{k}}(\eta)h^{s,*}_{\boldmathsymbol{k}^\prime}(\eta)\rangle \equiv \delta^{(3)}(\boldmathsymbol{k} - \boldmathsymbol{k}^\prime) \delta^{rs} \frac{2\pi^2}{k^3}\mathcal{P}^s_{h}(\eta,k),
\eea
where $s=(\times)$ or $(+)$ or $(\mathrm{sc})$.
At the end,  after a straightforward but rather long calculation one acquires that $\mathcal{P}_{h}(\eta,k)$ for the $(\times)$ and $(+)$ polarization states can be recast as ~\cite{Ananda:2006af,Baumann:2007zm,Kohri:2018awv,Espinosa:2018eve} 
\bea
\label{Tensor Power Spectrum}
\mathcal{P}^{(\times)\;\mathrm{or}\; (+)}_h(\eta,k) = 4\int_{0}^{\infty} \mathrm{d}v\int_{|1-v|}^{1+v}\mathrm{d}u \left[ \frac{4v^2 - (1+v^2-u^2)^2}{4uv}\right]^{2}I^2(u,v,x)\mathcal{P}_\Phi(kv)\mathcal{P}_\Phi(ku)\,,
\eea
whereas for the scalaron polarization one obtains that 
\bea
\label{Tensor Power Spectrum-scalaron}
\mathcal{P}^{(\mathrm{sc})}_h(\eta,k) = 8\int_{0}^{\infty} \mathrm{d}v\int_{|1-v|}^{1+v}\mathrm{d}u \left[ \frac{ (1+v^2-u^2)^2}{4uv}\right]^{2}I^{2}(u,v,x)\mathcal{P}_\Phi(kv)\mathcal{P}_\Phi(ku)\,.
\eea
 The two auxiliary variables $u$ and $v$ are defined as $u \equiv |\boldmathsymbol{k} - \boldmathsymbol{q}|/k$ and $v \equiv q/k$, and the kernel function $I(u,v,x)$ is given by
\bea
\label{I function}
I(u,v,x) = \int_{x_\mathrm{d}}^{x} \mathrm{d}\bar{x}\, \frac{a(\bar{x})}{a(x)}\, k\, G^s_{k}(x,\bar{x}) F_k(u,v,\bar{x}).
\eea
In the above expressions, $x=k\eta$ and we use the notation $F_{k}(u,v,\eta)\equiv  F(k ,|\boldmathsymbol{k}-\boldmathsymbol{q}|,\eta)$ since  the function $F(\boldmathsymbol{q},\boldmathsymbol{k-q},\eta)$ depends only on the modulus of its first two arguments. In the following, since we focus on second-order effects, we assume that the background evolution is close to that of $\Lambda\mathrm{CDM}$ scenario, and since  in the time period we are investigating the Universe is matter (i.e. PBH) dominated, we have  $w_\mathrm{tot} \simeq w_\mathrm{PBH}= 0$. Under these considerations, in a matter era, the Bardeen potential is, up to a decaying mode, constant in time, hence $T_\Phi = 1$ and from \Eq{F}, one gets that $F=10/3$. 

Finally, note also that the power spectrum of the PBH gravitational potential should be calculated at a reference initial time, which here is considered to be the PBH domination time.

\subsection{The gravitational wave energy desity spectrum}\label{subsec:rho_GW}

Since we have extracted  the power spectrum of the tensor perturbations, we can now calculate  the energy density associated to the SIGWs. We focus  only on subhorizon scales, in which one does not feel the curvature of spacetime and hence he can use a flat spacetime approximation. Consequently, after a straigthforward but lengthy calculation the GW energy density can be recast as \cite{Maggiore:1999vm}
\bea
\label{rho_GW effective}
 \rhoGW (\eta,\boldmathsymbol{x}) =  \frac{\Mp^2}{32 a^2}\, \overline{\left(\partial_\eta h_\mathrm{\alpha\beta}\partial_\eta h^\mathrm{\alpha\beta} +  \partial_{i} h_\mathrm{\alpha\beta}\partial^{i}h^\mathrm{\alpha\beta} \right)}\, ,
\eea
which is simply the sum of a kinetic term and a gradient term.  The overall bar stands for an oscillation averaging on sub-horizon scales, which is performed  to deduce only the envelope of the gravitational-wave spectrum. The GW spectral abundance is just the GW energy density per logarithmic comoving scale, i.e. 
\beq\label{Omega_GW}
\Omega_\mathrm{GW}(\eta,k) = \frac{1}{\bar{\rho}_\mathrm{tot}}\frac{\mathrm{d}\rho_\mathrm{GW}(\eta,k)}{\mathrm{d}\ln k}.
\eeq

Let us now focus on a matter-dominated era driven by PBHs, where $w=0$. Under these conditions, the transfer function $T_\mathrm{\Phi}$ is constant in time, and we normalise it to one at PBH domination time, namely $T_\mathrm{\Phi}(x_\mathrm{d})=1$. This forces the source term $S^s_\boldmathsymbol{k}$ to be constant in time and as a consequence at sub-horizon scales, where $k\gg \cal{H}$, from \Eq{Tensor Eq. of Motion} we acquire that $h^s_\boldmathsymbol{k}\simeq \frac{4S^s_\boldmathsymbol{k}}{k^2}$. Consequently, the tensor modes have a mild dependence on time and therefore the kinetic term in relation (\ref{rho_GW effective}) gives a negligible contribution to the GW energy density. Therefore, we straightforwardly obtain   that 
\beq\label{rho_GW_effective MD}
\begin{split}
\left\langle \rhoGW (\eta,\boldmathsymbol{x}) \right\rangle &  \simeq \left\langle \rho_{\mathrm{GW,grad}} (\eta,\boldmathsymbol{x}) \right\rangle = \sum_{s=+,\times,\mathrm{sc}}\frac{\Mp^2}{32a^2}\overline{\left\langle\left(\nabla h^{s}_\mathrm{\alpha\beta}\right)^2\right \rangle }
   \\ & =   \frac{\Mp^2}{32a^2 \left(2\pi\right)^3} \sum_{s=+,\times,\mathrm{sc}} \int\mathrm{d}^3\boldmathsymbol{k}_1 \int\mathrm{d}^3\boldmathsymbol{k}_2\,  k_1 k_2 \overline{  \left\langle h^{s}_{\boldmathsymbol{k}_1}(\eta)h^{s,*}_{\boldmathsymbol{k}_2}(\eta)\right\rangle} e^{i(\boldmathsymbol{k}_1-\boldmathsymbol{k}_2)\cdot \boldmathsymbol{x}}\,,
 \end{split}
\eeq
where the brackets stand for an ensemble average.
At the end, by combining \Eq{rho_GW_effective MD}, \Eq{Omega_GW} and \Eq{tesnor power spectrum definition} and taking into account from \Eq{Tensor Power Spectrum} that the $(\times)$ and $(+)$ polarization modes give an equal contribution, we find that 
\beq\label{Omega_GW_sub_horizon}
\Omega_\mathrm{GW}(\eta,k) \simeq  \frac{1}{\bar{\rho}_\mathrm{tot}}\frac{\mathrm{d}\rho_\mathrm{GW,grad}(\eta,k)}{\mathrm{d}\ln k} =  \frac{1}{96}\left(\frac{k}{\calH(\eta)}\right)^{2}\left[2\overline{\mathcal{P}^{(\times)}_h}(\eta,k)+\overline{\mathcal{P}^{(\mathrm{sc})}_h}(\eta,k)\right].
\eeq

\section{The case of Starobinsky $R^2$  gravity}
\label{ConstrSrarob}

In this section, by demanding that SIGWs are not overproduced at PBH evaporation time, firstly we derive constraints on the PBH abundances in the context of Starobinsky $R^2$ modified gravity with the mass scale $M$ taking its fiducial value $M=10^{-5}\Mp$. Afterwards, by treating $R^2$ as an illustrative $f(R)$ gravity case study theory, we treat its mass parameter $M$ as a free parameter and by avoiding again GW overproduction, we set constraints this time on $M$. These constraints on $M$ however should not be viewed as physical ones since $M$ is well fixed by the amplitude of the scalar perturbations through CMB probes. They are derived only to demonstrate that with the case study example of $R^2$ gravity, the portal of SIGWs associated to PBH Poisson fluctuations can be used as a novel probe to constrain the underlying gravity theory.

In our setup, we investigate and extract the SIGW spectrum produced during a cosmic era driven by PBHs. In order to achieve this  we treat the PBHs as a matter fluid, and thus with zero equation-of-state parameter, an approximation which is justifiable for scales larger than the PBH mean PBH separation scale where $k<k_\mathrm{UV}$ (see the discussion in subsection \ref{sec:PowerSpectrumPhiGR}).

\subsection{The theoretical parameters involved}
Before going into the investigation of the GW signal let us discuss the relevant theoretical parameters involved in the problem at hand. These parameters are actually the mass of the PBH $m_\mathrm{PBH}$, the initial PBH abundance at formation time $\Omega_\mathrm{PBH,f}$, and the dimensionless parameter $\alpha$ defined as the ratio of the Hubble parameter at PBH formation time over the energy scale parameter of Starobinsky (or $R^2$) gravity $M$
\beq\label{eq:a_definition}
\alpha\equiv H_\mathrm{f}/M.
\eeq

Regarding the PBH mass range we assume that the PBHs considered here are formed after the end of inflation and evaporate before BBN time. In particular, we extract a lower and an upper bound on the PBH mass, $m_\mathrm{PBH}$ by accounting for the current Planck upper bound on the tensor-to-scalar ratio for single-field slow-roll models of inflation, which gives  $\rho^{1/4}_\mathrm{inf}<10^{16}\mathrm{GeV}$ ~\cite{Planck:2018jri} as well conservative a lower bound on the reheating energy scale, i.e. $\rho^{1/4}_\mathrm{reh}> 4\mathrm{MeV}$~\cite{Kawasaki:1999na,Kawasaki:2000en,Hasegawa:2019jsa,Carr:2020gox}. Consequently, by requiring that $\rho_\mathrm{reh}\geq\rho_\mathrm{BBN}$ and considering the fact that the mass of a PBH is roughly equal to the mass inside the Hubble volume at PBH fomation time, $m_\mathrm{PBH}=4\pi\rho_\mathrm{f}H^{-3}_\mathrm{f}/3$, we can straightforwardly show that the relevant PBH mass range is given by 
\bea
\label{eq:domain:mPBHf}
10 \mathrm{g}< m_\mathrm{PBH}< 10^{9} \mathrm{g}\, ,
\eea
where moreover we have used the fact that the Hawking evaporation time of a black hole scales with the mass  $m_\mathrm{PBH}$ as $t_\mathrm{evap}=\frac{160}{\pi g_\ueff}\frac{m^3_\mathrm{PBH}}{\Mp^4}$ ~\cite{Hawking:1974rv}, where $g_\ueff$ is the effective number of relativistic degrees of freedom. In our numerical applications, we take $g_\ueff=100$ since it is the order of magnitude predicted by the Standard Model before the electroweak phase transition~\cite{Kolb:1990vq}.

Concerning now the range of $\Omega_\mathrm{PBH,f}$ in order to have a transient PBH domination era, this can be set by demanding that the PBH evaporation time $t_\mathrm{evap}$, is larger than the PBH domination time $t_\mathrm{d}$. In particular, knowing that during a radiation domination era $\Omega_\mathrm{PBH}=\rho_\mathrm{PBH}/\rho_\mathrm{d} \propto a^{-3}/a^{-4}\propto a$, then the PBHs dominate the energy budget of the Universe when $\Omega_\mathrm{PBH}=1$, from which we find that $a_\mathrm{d}=a_\mathrm{f}/\Omega_\mathrm{PBH,f}$. Thus, knowing that during radiation domination era $H\simeq 1/(2t)$, and demanding  that $t_\mathrm{evap}>t_\mathrm{d}$, we obtain  that 
\bea
\label{eq:domain:OmegaPBHf}
\Omega_\mathrm{PBH,f} >   10^{-15} \sqrt{\frac{g_\ueff}{100}} \frac{10^9\mathrm{g}}{m_\mathrm{PBH}}\, .
\eea

Finally, regarding the dimensionless parameter $\alpha$, knowing that in Starobinsky gravity $M\sim H_\mathrm{inf}$,
and assuming as mentioned above that PBHs are formed after inflation, i.e. $ H_\mathrm{inf}\geq H_\mathrm{f}$, we get that $M\geq H_\mathrm{f}$. In addition, we know that in Starobinsky-like inflationary models $M<M_\mathrm{max}\equiv 10^{-5}\Mp$ in order to be compatible with the amplitude of curvature power spectrum from CMB observations. Consequently, for our considerations we have that $H_\mathrm{f}\leq M\leq M_\mathrm{max}$ and the relevant range for $\alpha$ can be recast as
\bea
\label{eq:domain:alpha}
\frac{H_\mathrm{f}}{M_\mathrm{max}} \leq \alpha \leq 1\, .
\eea

In the limit $\alpha\rightarrow 0 \Leftrightarrow M\rightarrow \infty$ one recovers GR. However, given the fact that we constrain our analysis to regimes where $M\leq 10^{-5}\Mp$, the GR limit $\alpha\rightarrow 0$ is not included here.

At this point, we need to stress that given the fact that the energy scale $M$ is more or less the energy scale at the end of inflation, the regime where $\alpha \sim 1$, or equivalently $M\sim H_\mathrm{f}$, corresponds to a regime where PBHs are created right after the end of inflation considering instantaneous reheating. This assumption of instantaneous reheating is a simplistic one but rather reasonable given the fact that in the following we aim to illustrate with the case study of $R^2$ gravity how one can set constraints on the theoretical parameters of the underlying $f(R)$ gravity theory using the novel probe of the SIGWs associated to PBH Poisson fluctuations rather than extract precise constraints on $M$.

\subsection{Gravitational waves from an era driven by primordial black holes}
Having introduced in the previous subsection the relevant parameters involved we derive here the GW spectrum during an era of PBH domination. To do so, the first step is to calculate the kernel function $I(u,v,x)$ defined in \Eq{I function}. Since we are in a matter (i.e. PBHs) dominated era, namely with $w=0$,   in the subhorizon limit, i.e. $x\gg 1$, $I(u,v,x)$ reads as (see \App{sec:I(v,u,x)}) 
\beq\label{I_2_approx}
I^2(x) = \frac{100}{9}\times
\begin{cases}
1 \mathrm{\;if\;s=(\times),(+)}\\
\frac{k^4}{M^4}  \mathrm{\;if\;s=(\mathrm{sc})}
\end{cases}.
\eeq
As one may notice from the expression (\ref{I_2_approx}), we have a suppression factor of the order $k^4/M^4$, which suppresses the scalaron contribution. One expects the highest contribution of this factor in the region close to the UV cut-off scale and the regimes where $M$ takes its minimum value, namely $M=M_\mathrm{min}=H_\mathrm{f}$. In particular, when $k=k_\mathrm{UV}$ and $M=H_\mathrm{f}$ one gets that $$I^2_\mathrm{(sc)}(x) =\frac{100}{9} \Omega_\mathrm{PBH,f}^{4/3}<1<I^2_\mathrm{(+) or (\times)}(x) = \frac{100}{9}.$$\footnote{For the derivation of the expression of $I^2_\mathrm{(sc)}(x)$ and $I^2_\mathrm{(+) or (\times)}(x)$ see \App{sec:I(v,u,x)}.}
since $\Omega_\mathrm{PBH,f}<1$. As a result, one anticipates that the scalaron contribution should be negligible with respect to the contributions from the $(+)$ and $(\times)$ polarisations. This can be seen from the right panel of  \Fig{fig:GW_spectrum_at_evap_different_alpha} where we see that for $m_\mathrm{PBH}=10^3\mathrm{g}$, $\Omega_\mathrm{PBH,f}=10^{-3}$ and $M=H_\mathrm{f}$ the scalaron contribution to the GW signal is indeed the subdominant one.

Under this approximation, neglecting the scalaron contribution, the GW spectrum (\ref{Omega_GW_sub_horizon}) can be recast in the following form:
\bea
\label{Omega_2nd_order_MD_newtonian_gauge_UV_cutoff}
 \OmegaGW (\eta,k)  = \frac{4}{75\pi^2}\left(\frac{k}{aH}\right)^{2}\left(\frac{k}{k_\mathrm{UV}}\right)^6  \, 
{\cal F}\left(y,  \Omega_\mathrm{PBH,f},\alpha \right),
\eea
where
{\small{
\beq
\label{curly F}
{\cal F}(y,  \Omega_\mathrm{PBH,f},\alpha )=\int_{0}^{\Lambda_\mathrm{UV}} \mathrm{d}v\int_{|1-v|}^{\min(\Lambda_\mathrm{UV},1+v)}\mathrm{d}u\left[ \frac{4v^2 - (1+v^2-u^2)^2}{4\left(3+\frac{2\Xi(\alpha,\Omega_\mathrm{PBH,f})}{5}y^2v^2\right)\left(3+\frac{2\Xi(\alpha,\Omega_\mathrm{PBH,f})}{5}y^2u^2\right)}\right]^{2} {uv}, 
\eeq}}
with $y=k/(a_\mathrm{d}H_\mathrm{d})$. $\Lambda_\mathrm{UV}$ is the upper bound of the integral in $v$ due to the UV cut-off scale, discussed in subsection  \ref{sec:PowerSpectrumPhiGR}, and is defined as  \cite{Papanikolaou:2020qtd}
\beq\label{Lambda_UV}
\Lambda_\mathrm{UV} = \frac{k_\mathrm{UV}}{k}.
\eeq
Finally, the function $\Xi(\alpha,\Omega_\mathrm{PBH,f})$  is defined as 
\beq\label{Xi(alpha)}
\Xi(\alpha,\Omega_\mathrm{PBH,f}) = \frac{F(a_\mathrm{d})}{\xi(\alpha,\Omega_\mathrm{PBH,f})}\left[\frac{1+ 3 \frac{k^2}{a_\mathrm{d}^2} \frac{F_{,R}(a_\mathrm{d})}{F(a_\mathrm{d})}}{1+ 2 \frac{k^2}{a_\mathrm{d}^2}\frac{F_{,R}(a_\mathrm{d})}{F(a_\mathrm{d})}}\right]  \simeq  \frac{1}{\xi(\alpha,\Omega_\mathrm{PBH,f})},
\eeq
where  $\Xi(\alpha,\Omega_\mathrm{PBH,f}) \simeq 1/\xi(\alpha,\Omega_\mathrm{PBH,f})$
since as as we have verified numerically  $F(a_\mathrm{d}) = 1 +\alpha^2\Omega^2_\mathrm{PBH,f}\sim 1$ and  $\left(1+ 3 \frac{k^2}{a^2} \frac{F_{,R}(a_\mathrm{d})}{F(a_\mathrm{d})}\right)/\left(1+ 2 \frac{k^2}{a^2}\frac{F_{,R}(a_\mathrm{d})}{F(a_\mathrm{d})}\right) \sim 1$. Note that we have dropped the argument $a_\mathrm{d}$ from $\xi$ in order not to have a heavy notation and we will keep this convention throughout the paper.

\begin{figure}[ht]
\begin{center}
\includegraphics[scale=.95]
{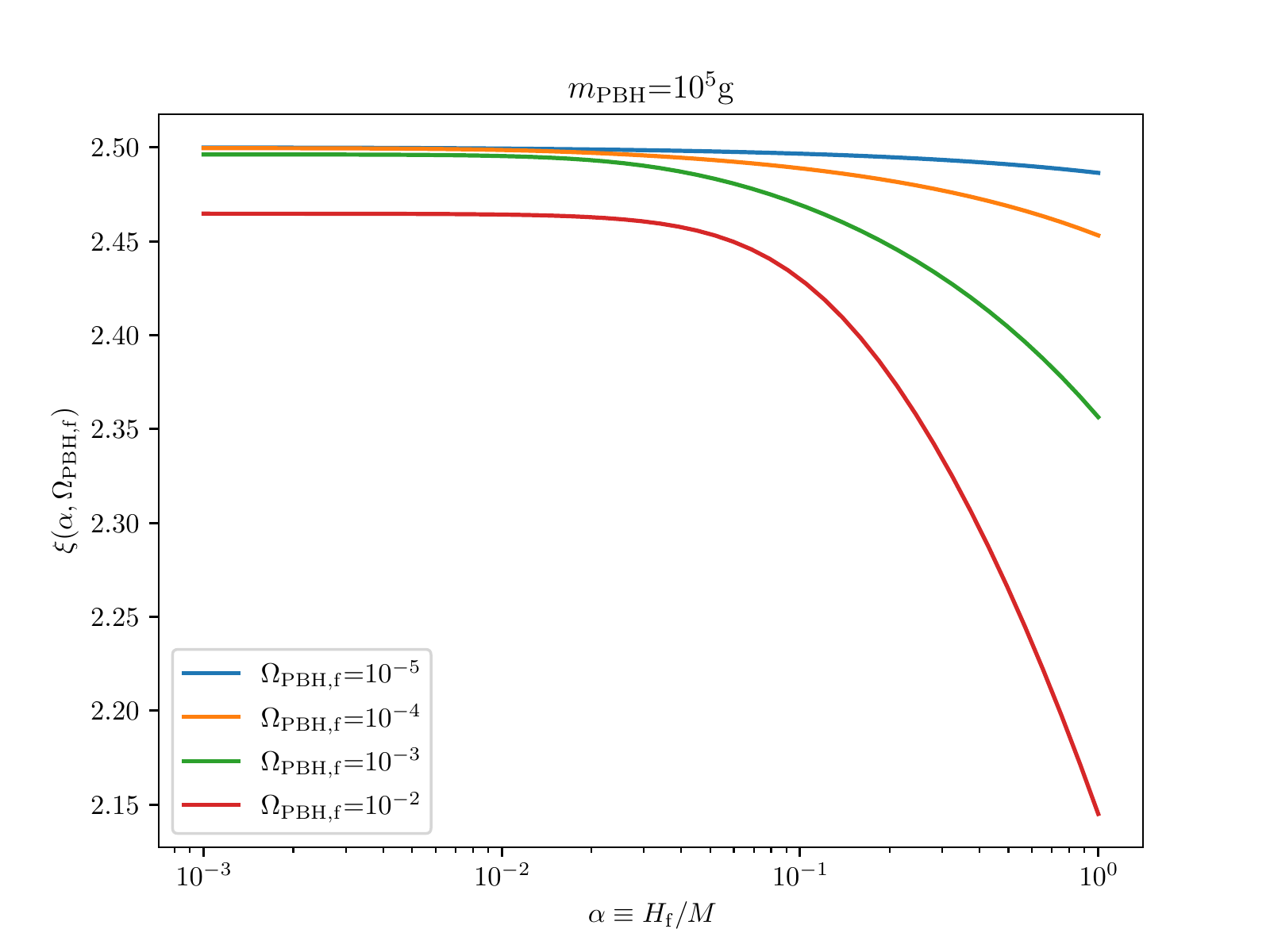}
\end{center}
\caption{{\it{ The ratio of the PBH density contrast computed at PBH domination time over the PBH density contrast at PBH formation  $\xi(\alpha,\Omega_\mathrm{PBH,f})$ given in   (\ref{eq:xi_definition}),  as a function of $\alpha$, for fixed
 $m_\mathrm{PBH}=10^5\mathrm{g}$ and  for various values of $\Omega_\mathrm{PBH,f}$.}}}
\label{fig:xi_alpha}
\end{figure}

\begin{figure}[ht]
\begin{center}
\includegraphics[width=0.496\textwidth, clip=true]
{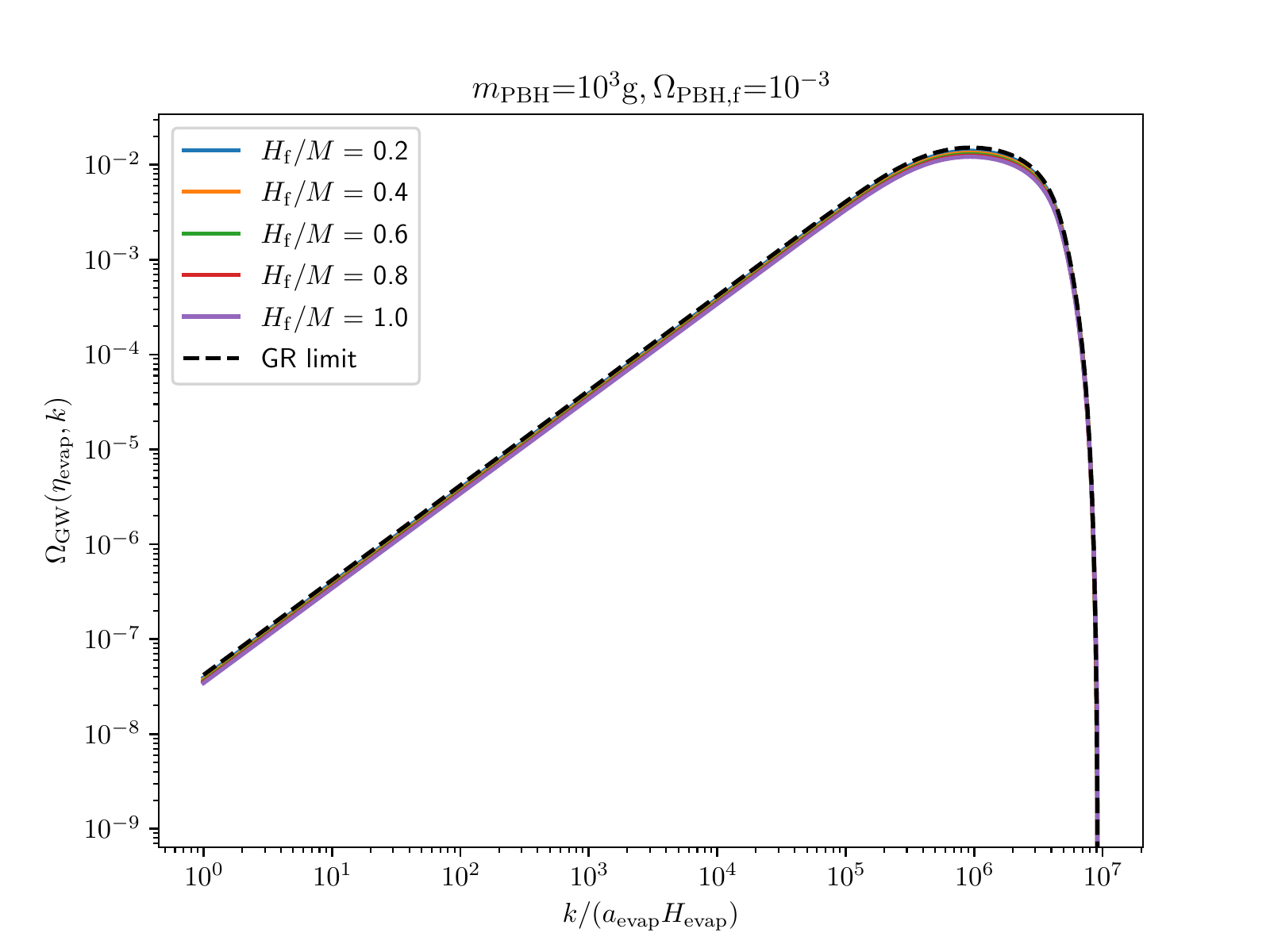}
\includegraphics[width=0.496\textwidth, clip=true]{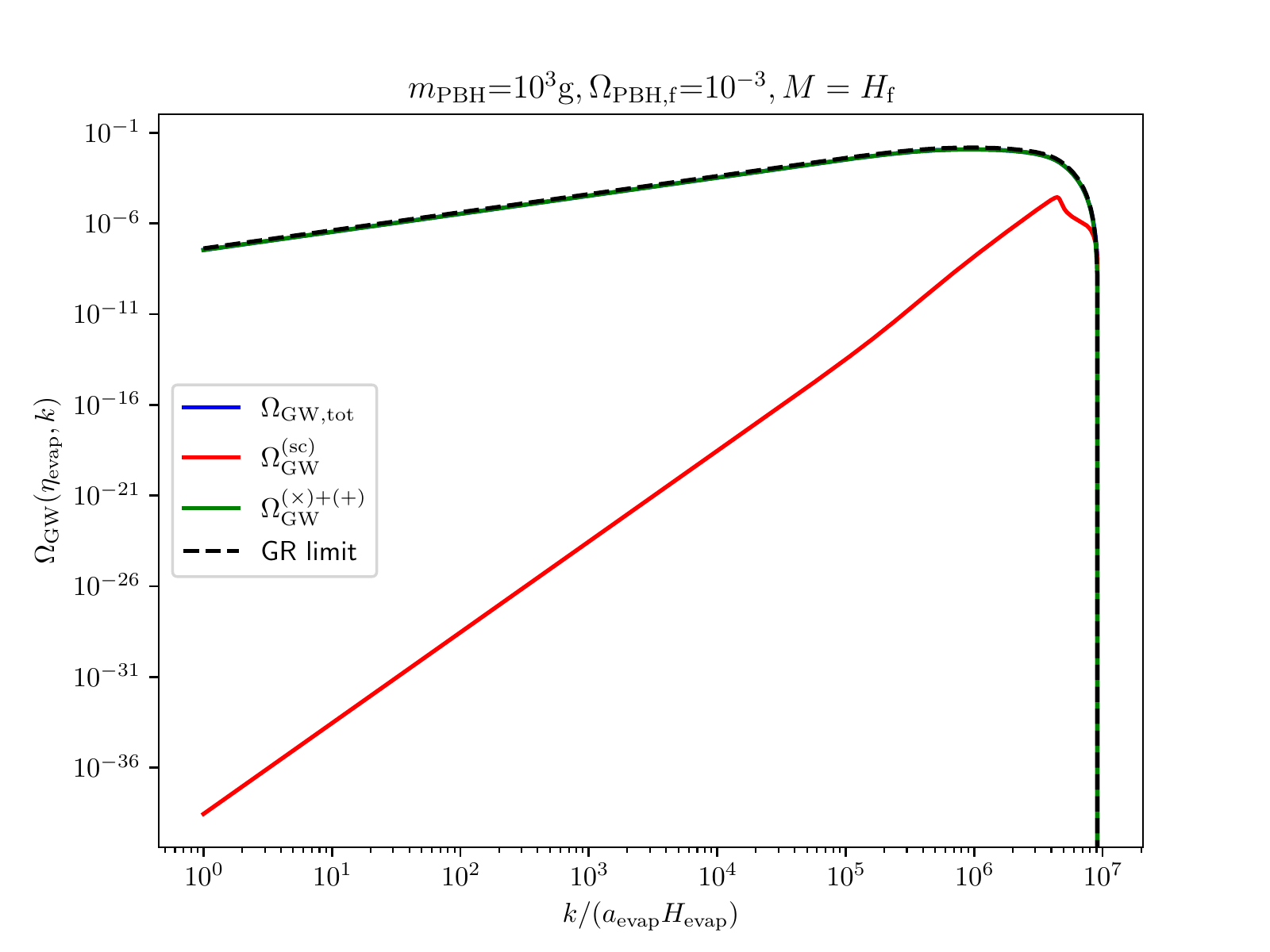}
\end{center}
\caption{{\it{Left Panel: The GW spectral abundance $\Omega_\mathrm{GW}(\eta_\mathrm{evap},k)$ at PBH evaporation time, for various values of the parameter $\alpha= H_\mathrm{f}/M$, in the case where $m_\mathrm{PBH}=10^5\mathrm{g}$ and $\Omega_\mathrm{PBH,f}=5\times 10^{-5}$. The dashed black curve represents the GR limit. Right Panel: The contributions from the scalaron and $(+) + (\times)$ to the GW spectral abundance $\Omega_\mathrm{GW}(\eta_\mathrm{evap},k)$ at PBH evaporation time, in the case where  $m_\mathrm{PBH}=10^5\mathrm{g}$ and $\Omega_\mathrm{PBH,f}=\times 10^{-5}$ and $M=10^{-5}\Mp$. The dashed black curve represents the GR limit. }}}
\label{fig:GW_spectrum_at_evap_different_alpha}
\end{figure}

In \Fig{fig:xi_alpha} we depict the function $\xi(\alpha,\Omega_\mathrm{PBH,f})$ as a function of $\alpha$, taking different values of $\Omega_\mathrm{PBH,f}$.
As we observe, $\xi(\alpha,\Omega_\mathrm{PBH,f})$ is a decreasing function of $\alpha$, with a plateau behaviour for small values of $\alpha$. For relatively small $\Omega_\mathrm{PBH}$ values we can also infer that $\xi(\alpha,\Omega_\mathrm{PBH,f})$ depends sligtly on $\Omega_\mathrm{PBH,f}$.

Consequently, having calculated $\xi(\alpha,\Omega_\mathrm{PBH,f})$, we can insert it in  expression (\ref{Omega_2nd_order_MD_newtonian_gauge_UV_cutoff}) and extract the GW spectrum. In the left panel of \Fig{fig:GW_spectrum_at_evap_different_alpha} we show the GW spectral abundance at PBH evaporation time, $\Omega_\mathrm{GW}(\eta_\mathrm{evap},k)$, namely at the end of the PBH-dominated era, for different values of the parameter $\alpha= H_\mathrm{f}/M$. As one may see, as $\alpha$ increases we have a departure from the GR limit which can be clearly observed in the regime where $\alpha \sim 1$ or equivalently when $M\sim H_\mathrm{f}$. In particular, the increase of $\alpha$ decreases the GW signal, due to the fact that for  fixed $\Omega_\mathrm{PBH,f}$, $\xi(\alpha,\Omega_\mathrm{PBH,f})$ is a decreasing function
of $\alpha$, as it can be seen from \Fig{fig:xi_alpha}. In terms now of the mass parameter $M$, an increase in $M$ is equivalent with an increase in the amplitude of the GW signal. Concerning now the contribution of the different polarisation states to the amplitude of GWs we find a negligible contribution of the scalaron polarisation, a fact which leaves the shape of the GW spectrum the same as that of GR. This behavior can be confirmed by right panel of \Fig{fig:GW_spectrum_at_evap_different_alpha}. 

\subsection{Gravitational wave backreaction constraints}
Interestingly enough, according to the above analysis we deduce that for some values of the involved parameters  one is met with an overproduction of gravitational waves at PBH evaporation time, which is something unphysical. This GW overproduction issue seems rather intriguing since one would expect that the energy density of gravitational waves generated by PBHs inhomogeneties decays like radiation as $a^{-4}$, i.e. faster than the energy density of PBHs themselves which decays like matter as  $a^{-3}$. This is true in the case where GWs decay as free waves. However, in our case we study the SIGW production during an early PBH domination era with the source of the induced GWs, namely \Eq{eq:Source:def} not being zero. Under these conditions, the tensor perturbations are not decoupled from the scalar ones and the GWs are not freely propagating. One then expects a continuous production of GWs up to the time where the source term (\ref{eq:Source:def}) has sufficiently decayed, namely after the PBH evaporation time. After this time, GWs evolve as radiation with $\rho_\mathrm{GW}\sim a^{-4}$. Therefore, in order to avoid this GW backreaction issue we demand that $\Omega_\mathrm{GW,tot}(\eta_\mathrm{evap})<1$. Hence, this condition will lead to bounds for the relevant parameters of the problem at hand. 
\subsubsection{Constraints on the primordial black hole abundance}
Following the aforementioned discussion we extract below analytical constraints on the initial abundance of PBHs $\Omega_\mathrm{PBH,f}$ as a function of the PBH mass $m_\mathrm{PBH}$ and the mass parameter $M$ of $R^2$ gravity. In order to achieve this, one can expand $\mathcal{F}$ in the regimes $y\ll 1$ and $y\gg 1$. Following the procedure described in   Appendix B of  \cite{Papanikolaou:2020qtd} one obtains that \footnote{The full expression for $\mathcal{F}(y, \Omega_\mathrm{PBH,f} )$ independently of $\Omega_\mathrm{PBH,f} $ is given by 
\beq
\begin{aligned}
\mathcal{F}(y\ll 1, & \Omega_\mathrm{PBH,f} ) =  \frac{500\xi^{7/2}(\alpha,\Omega_\mathrm{PBH,f})}{576}\Biggl[\sqrt{30}\mathrm{ArcTan}\left(\sqrt{\frac{2}{15\xi(\alpha,\Omega_\mathrm{PBH,f})}}\frac{1}{\Omega^{2/3}_\mathrm{PBH,f}}\right)  \\ &
\!\!\!
- \frac{6}{\sqrt{\xi(\alpha,\Omega_\mathrm{PBH,f})}}\frac{\left(1125\Omega^{10/3}_\mathrm{PBH,f} + 44\xi^{-2}(\alpha,\Omega_\mathrm{PBH,f})\Omega^{2/3}_\mathrm{PBH,f} + 400\Omega^2_\mathrm{PBH,f}/\xi(\alpha,\Omega_\mathrm{PBH,f}) \right)}{\left(15\Omega^{4/3}_\mathrm{PBH,f}+\frac{2}{\xi(\alpha,\Omega_\mathrm{PBH,f})}\right)^2} \Biggr]
\end{aligned}.
\eeq
} 

\bea
\label{eq:calF:approx}
\mathcal{F}(y, \Omega_\mathrm{PBH,f} )\simeq 
\begin{cases}
\frac{125}{48}\sqrt{\frac{5}{6}}\frac{\pi\xi^{7/2}(\alpha,\Omega_\mathrm{PBH,f})}{y^7} \mathrm{\;for\;}y\ll 1 \mathrm{\;and\;}   \Omega_\mathrm{PBH,f} \ll 1 \\ 
 \frac{625\pi^2\xi^4(\alpha,\Omega_\mathrm{PBH,f}) }{128y^8}\mathrm{\ for\ }y\gg 1
\end{cases} .
\eea
 Then, inserting the above expression into (\ref{Omega_2nd_order_MD_newtonian_gauge_UV_cutoff}) we acquire
\begin{eqnarray}
\label{Omega_2nd_order_MD_newtonian_gauge_analytical_approximation - k<<a_dH_d}
\Omega_\mathrm{GW}(\eta_\mathrm{evap},k\ll \calH_\mathrm{d}) 
 & \simeq &\frac{8\sqrt{2}}{3}\frac{\xi^{7/2}(\alpha,\Omega_\mathrm{PBH,f})}{\pi }\!\left(\frac{g_\mathrm{eff}}{100}\right)^{-2/3}\! \frac{k}{\calH_\ud}\left(\frac{ m_\mathrm{PBH} }{\Mp}\right)^{4/3}\!\!\Omega^{16/3}_\mathrm{PBH,f}\,  ,
 \quad\quad
 \\
\label{Omega_2nd_order_MD_newtonian_gauge_analytical_approximation - k>>a_dH_d}
\Omega_\mathrm{GW}(\eta_\mathrm{evap},k\gg \calH_\mathrm{d}) 
& \simeq & 50\left(\frac{3}{5}\right)^{3/2}\xi^4(\alpha,\Omega_\mathrm{PBH,f})\!\left(\frac{g_\mathrm{eff}}{100}\right)^{-2/3}\! \left(\frac{ m_\mathrm{PBH} }{\Mp}\right)^{4/3}\!\!\Omega^{16/3}_\mathrm{PBH,f}\, .
\end{eqnarray}

Finally, by integrating over $\ln k$ we obtain the total amount of GWs produced during the PBH domination era, namely
\beq
\label{eq:OmegaGW:tot}
\Omega_\mathrm{GW,tot}(\eta_\mathrm{evap}) = \int\mathrm{d}\ln k \  \OmegaGW (\eta_\mathrm{evap},k).
\eeq
Specifically, by replacing (\ref{Omega_2nd_order_MD_newtonian_gauge_analytical_approximation - k<<a_dH_d}) and (\ref{Omega_2nd_order_MD_newtonian_gauge_analytical_approximation - k>>a_dH_d}) into (\ref{eq:OmegaGW:tot}), $\Omega_\mathrm{GW,tot}(\eta_\mathrm{evap})$ is written as
\bea
\Omega_\mathrm{GW,tot}(\eta_\mathrm{evap})=\mu\left[\kappa-\ln(\Omega_\mathrm{PBH,f})\right]\Omega^{16/3}_\mathrm{PBH,f}\, ,
\eea
with 
\bea
\mu= 20\xi^{7/2}(\alpha,\Omega_\mathrm{PBH,f})\left(\frac{3}{5}\right)^{1/2}\left(\frac{g_\ueff}{100}\right)^{-2/3} \left(\frac{m_\mathrm{PBH}}{\Mp}\right)^{4/3}
\eea
 and 
\bea
\kappa = \frac{2\sqrt{2}}{9}\frac{1}{\pi\sqrt{\xi(\alpha,\Omega_\mathrm{PBH,f})}}+\frac{3}{2}\ln2\, .
\eea
As a last step, let as extract the bounds for the parameters $m_\mathrm{PBH}$, $\Omega_\mathrm{PBH,f}$ and $\alpha$. To do so, we need to solve the equation $\Omega_\mathrm{GW,tot}(\eta_\mathrm{evap})= 1$. This equation can be solved in terms of the Lambert function~\cite{Olver:2010:NHM:1830479:Lambert}, obtaining
\bea
\Omega_\mathrm{PBH,f}^\umax=\left[ -\frac{3 \mu}{16}W_{-1}\left(-\frac{16}{3\mu}\ee^{-\frac{16\kappa}{3}}\right)\right]^{-3/16},
\eea
where $W_{-1}$ is the ``$-1$''-branch of the Lambert function. Given the fact that $m_\mathrm{PBH}>10\mathrm{g}$ [see \Eq{eq:domain:mPBHf}], we find that $\mu\gg 1$, while $\kappa$ is of order one. Consequently, the argument of the Lambert function is close to zero, and in this regime it can be approximated by a logarithmic function, i.e. $W_{-1}\left(-\frac{16}{3\mu}\ee^{-\frac{16\kappa}{3}}\right) \simeq |\ln\left(-\frac{16}{3\mu}e^{-\frac{16\kappa}{3}}\right)|$. Now taking  into account the mild dependence of the logarithmic function on its argument, for our numerical purposes we will choose a central value in PBH mass range, namely $m_\mathrm{PBH}= 10^5\mathrm{g}$, and we will consider the logarithm as constant. Concerning the value of $\xi(\alpha)$, given the fact that for $\Omega_\mathrm{PBH,f}\leq 0.01$ it varies between $2.2$ and $2.5$ (see \Fig{fig:xi_alpha}) we will take it equal to $2.4$. 
\begin{figure}[h!]
\begin{center}
 \includegraphics[scale=.85]{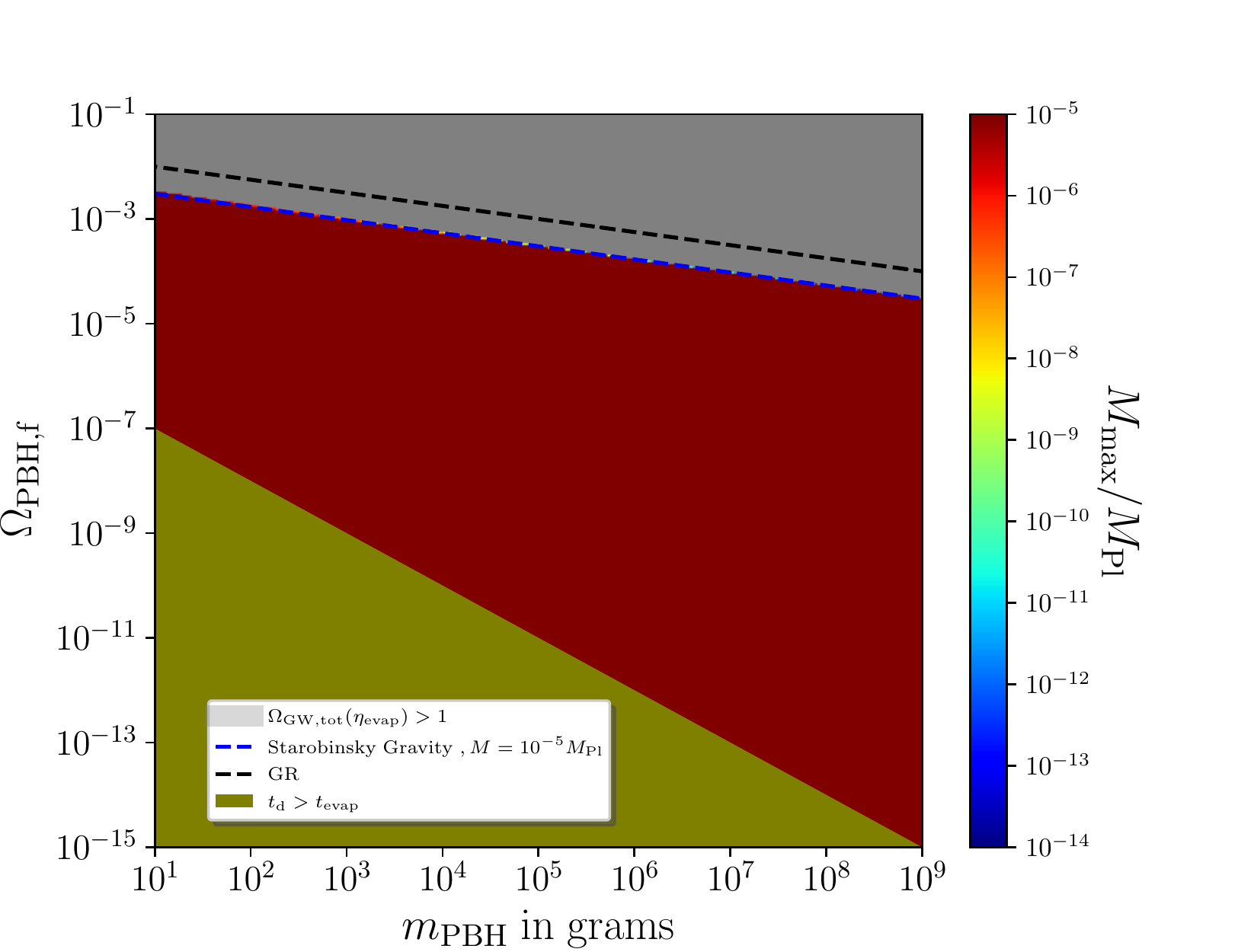}
\end{center}
\caption{{\it{The maximum of the mass parameter $M$ of   Starobinsky gravity in terms of the reduced Planck mass $\Mp$ (color bar axis) as a function of the PBH mass $m_\mathrm{PBH}$ (x axis) and the initial PBH abundance $\Omega_\mathrm{PBH,f}$ ($y$ axis). The values of $m_\mathrm{PBH}$ are chosen such that PBHs form after inflation and evaporate before Big Bang Nucleosynthesis, see (\ref{eq:domain:mPBHf}), whereas the displayed values of $\Omega_\mathrm{PBH,f}$ correspond to regimes where PBHs dominate the energy budget of the Universe for a transient period, see (\ref{eq:domain:OmegaPBHf}). The grey region corresponds to regimes where gravitational waves are overproduced at PBH evaporation time whereas the olive region stand for regimes where PBHs dominate after the completion of their evaporation process. The black dashed line corresponds to the GR upper bound on $\Omega_\mathrm{PBH,f}$ where the blue dashed line stands for the respective upper bound within Starobinsky gravity with the fiducial value for the mass parameter $M=10^{-5}\Mp$. For the numerical applications we have used $g_\ueff=100$.}}}
\label{fig:M_constraints}
\end{figure}

At the end, we straightforwardly obtain that 
\bea\label{OmegaPBHf_upper_bound}
\Omega_\mathrm{PBH,f} \leq 10^{-4}\left(\frac{10^9\mathrm{g}}{m_\mathrm{PBH}}\right)^{1/4}\frac{1}{\xi^{21/32}(M,\Omega_\mathrm{PBH,f})},
\eea

where $\xi(M,\Omega_\mathrm{PBH,f})$ is expressed in terms of the mass parameter of Starobinsky gravity. Choosing now the fiducial value of $M=10^{-5}\Mp$ as dictated by the CMB observations on the amplitude of the curvature power spectrum and exploiting the fact that for $M=10^{-5}\Mp$, $\xi(M,\Omega_\mathrm{PBH,f})$ has a very mild dependence on $\Omega_\mathrm{PBH,f}$ - $\xi(M=10^{-5}\Mp,\Omega_\mathrm{PBH,f}) \sim 2.5$ for every value of $\Omega_\mathrm{PBH,f}$ - we find that the upper constraint on $\Omega_\mathrm{PBH,f}$ reads as
\bea\label{OmegaPBHf_upper_bound_Starobinsky}
\Omega_\mathrm{PBH,f} \leq 5.5 \times 10^{-5}\left(\frac{10^9\mathrm{g}}{m_\mathrm{PBH}}\right)^{1/4}.
\eea
This upper bound on $\Omega_\mathrm{PBH,f}$ is depicted with the dotted blue line in \Fig{fig:M_constraints}. As it was checked, it is $45\%$ reduced compared to the respective upper bound within GR which reads as~\cite{Papanikolaou:2020qtd} $\Omega_\mathrm{PBH,f} \leq 10^{-4}\left(\frac{10^9\mathrm{g}}{m_\mathrm{PBH}}\right)^{1/4}$. Thus, one finds that despite the fact that the corrections from the $R^2$ term are very small at the level of the background and perturbations, as it can be seen already from the left panel of Fig. 2, we find almost an order of magnitude tighter constraints on $\Omega_\mathrm{PBH,f}$ compared to GR. This result has important consequences at the level of the detectability of the SIGW signal associated to PBH Poisson fluctuations since, as we can see from \Eq{Omega_2nd_order_MD_newtonian_gauge_analytical_approximation - k<<a_dH_d} and \Eq{Omega_2nd_order_MD_newtonian_gauge_analytical_approximation - k>>a_dH_d}, given the mild dependence of $\xi(\Omega_\mathrm{PBH,f},M)$ on $\Omega_\mathrm{PBH,f}$, the amplitude of the signal scales as $\Omega^{16/3}_\mathrm{PBH,f}$ a scaling which also holds in GR.

\subsubsection{Constraints on the $f(R)$ gravity model at hand}
In our previous analysis we showed how one can constrain the PBH abundances by avoiding a GW overproduction issue. Conversely if one fixes the initial PBH abundance $\Omega_\mathrm{PBH,f}$ and their mass $m_\mathrm{PBH}$, one can translate the GW backreaction constraints to constraints on the underlying $f(R)$ gravity theory. Here we consider as an illustrative example, the case of Starobinsky $R^2$ gravity given that it is the simplest monoparametric extension of GR within the class of $f(R)$ gravity theories. In this sense, we will ignore the fact that the mass parameter $M$ is fixed by CMB observations to the value $M=10^{-5}\Mp$ and we will treat it as a free parameter.

Under these considerations, as we can see from the left panel of \Fig{fig:GW_spectrum_at_evap_different_alpha}, if one fixes $\Omega_\mathrm{PBH,f}$ and $m_\mathrm{PBH}$ by increasing the mass parameter $M$ or equivalently by decreasing the parameter $\alpha=H_\mathrm{f}/M$, the amplitude of SIGWs is increasing as well signalling that one can set an upper bound constraint on the mass parameter $M$. To do so, we solved numerically the equation $\Omega_\mathrm{GW,tot}(\eta_\mathrm{evap})=1$ and found the upper bound $M_\mathrm{max}$ on $M$ as a function of $m_\mathrm{PBH}$ and $\Omega_\mathrm{PBH,f}$. In \Fig{fig:M_constraints} we show this upper bound constraint on $M$ in the color bar axis. The lower left triangular region in ``olive" stands for the region in the parameter space  $(m_\mathrm{PBH},\Omega_\mathrm{PBH,f})$  where PBHs dominate the Universe energy content after their evaporation, hence it not of special interest. The upper grey region with large values of $\Omega_\mathrm{PBH,f}$ corresponds to regimes where GWs are overproduced during the PBH dominated era, so it is excluded. 

The interesting region which permits an early PBH dominated era not presenting a GW overproduction issue is the intermediate one, where we show in the lateral color bar axis the upper bound on the mass scale $M$. As expected, for the majority of the parameter space $(m_\mathrm{PBH},\Omega_\mathrm{PBH,f})$ $M_\mathrm{max}$
is found to be equal to $10^{-5}\Mp$ which is the fiducial value of $M$ as obtained by CMB observations. However, there is an interesting region between the ``bordeaux" region of $M_\mathrm{max}=10^{-5}\Mp$ and the grey region of GW overproduction where the upper bound $M_\mathrm{max}$ becomes smaller than $10^{-5}\Mp$ reaching very small values up to $10^{-14}\Mp$. This behavior can be explained from the fact that as in this region whose location is described by \Eq{OmegaPBHf_upper_bound_Starobinsky} $\Omega_\mathrm{PBH,f}$ takes its greatest value more or less between $10^{-4}$ up to $10^{-2}$ where one expects a high amplitude of GWs. To re-compensate therefore for this increased amplitude of GWs one should lower the mass scale $M$ since as we show in \Fig{fig:GW_spectrum_at_evap_different_alpha} the amplitude of GWs is an increasing function of $M$.

At this point, we should highlight that the upper bounds in $M$ should not be interpreted as physical since $M$ is very well fixed by the amplitude of the scalar perturbations as measured by CMB probes \footnote{In particular, the very low $M_\mathrm{max}$ regions are rather questionable since it is not easy to achieve scalaron decay quite early and thus one should account for very strong restrictions of $M$ in these regimes.}. As we already mentioned above, the choice of $R^2$ should rather be regarded as an illustrative example which demonstrates the fact that the SIGW portal associated to PBH Poisson fluctuations can serve as a novel probe to constrain alternative gravity theories.
\section{Conclusions}
\label{Conclusions}

Primordial black holes are of great significance, since they may constitute a part or all of the dark matter sector, they may provide an explanation for the large-scale 
structure formation through Poisson fluctuations, and moreover they can offer 
the seeds for the progenitors of the black-hole merging events as well as for 
the supermassive black holes formation. Their effect on the GW 
background signals, and in particular the second-order GWs induced by the gravitational potential of Poisson-distributed PBHs, has been studied only in the 
framework of general relativity. Hence, in this work we extended the analysis of the
literature in the case of  $f(R)$ gravity. In order to illustrate the effect of $f(R)$ gravity theory, we worked with the Starobinsky $R^2$ gravity, which constitutes the simplest monoparametric generalisation beyond GR within $f(R)$ theories as well as one of the most favored inflationary models from the observational side. However, our formalism is applicable for every model in the context of $f(R)$ gravity.

Firstly, we  calculated the effect of $f(R)$ modification on the PBH gravitational potential power spectrum and we extracted the associated SIGW spectrum during an era driven by ultralight PBHs ($m_\mathrm{PBH}<10^{9}\mathrm{g}$), which evaporate before BBN. In particular, we found its dependence on the relevant parameters involved, namely the PBH mass $m_\mathrm{PBH}$, the initial PBH abundance at formation time $\Omega_\mathrm{PBH,f}$, and the mass parameter of the $R^2$ gravity $M$ by accounting as well for the three polarization states of GWs in $f(R)$ gravity, namely the $(\times)$, the $(+)$ and the scalaron one. 

Concerning the contribution of the different polarisation states to the amplitude of GWs, we found a negligible contribution of the scalaron polarisation, a fact which left the shape of the GW spectrum the same as that of GR [See the right panel of \Fig{fig:GW_spectrum_at_evap_different_alpha}]. The only difference with respect to GR was observed at the level of the amplitude of GWs. In particular, a decrease of the mass parameter $M$ leads to a decrease of the GW amplitude which becomes distinguishable from the GW amplitude within GR in the region where $M\sim H_\mathrm{f}$, where $H_\mathrm{f}$ is the Hubble parameter at the PBH formation time [See the left panel of \Fig{fig:GW_spectrum_at_evap_different_alpha}].

Interestingly, in some region of our parameter space ($m_\mathrm{PBH}$, $\Omega_\mathrm{PBH,f}$, $M$) we found regimes where the overall energy density of the induced GWs at PBH evaporation time becomes greater than the total energy density of the Universe,   which is unphysical and thus needs to be avoided. Thus, in order to avoid this GW backreaction problem we demanded that the overall energy density contribution of the GWs at evaporation time is less than one, $\Omega_\mathrm{GW,tot}(\eta_\mathrm{evap})<1$. This condition allowed us to extract an upper bound on  $\Omega_\mathrm{PBH,f}$ as a function of the PBH mass and the mass parameter $M$. Intriguingly, this upper bound is the respective GR bound screened by a function of $M$ and $\Omega_\mathrm{PBH,f}$, namely
\bea\label{OmegaPBHf_upper_bound_conclusions}
\Omega_\mathrm{PBH,f} \leq 10^{-4}\left(\frac{10^9\mathrm{g}}{m_\mathrm{PBH}}\right)^{1/4}\frac{1}{\xi^{21/32}(M,\Omega_\mathrm{PBH,f})},
\eea

Given the above inequality condition, on the one hand, by fixing the mass parameter $M$ of $R^2$ gravity to its fiducial value $M=10^{-5}\Mp$ as imposed by CMB observations and exploiting the mild dependence of $\xi(M=10^{-5}\Mp,\Omega_\mathrm{PBH,f})$ on $\Omega_\mathrm{PBH,f}$\footnote{As it was found numerically $\xi(M=10^{-5}\Mp,\Omega_\mathrm{PBH,f})$ varies between the values $2.5$ and $2.47$ within the range for $\Omega_\mathrm{PBH,f}\in[10^{-15},10^{-1}]$. Thus, for our numerical purposes we take $\xi(M=10^{-5}\Mp,\Omega_\mathrm{PBH,f})=2.5$. } we found that 
\bea\label{OmegaPBHf_upper_bound_Starobinsky_conclusions}
\Omega_\mathrm{PBH,f} \leq 5.5\times 10^{-5}\left(\frac{10^9\mathrm{g}}{m_\mathrm{PBH}}\right)^{1/4},
\eea
which gives an upper bound $\Omega_\mathrm{PBH,f}$ $45\%$ tighter than that in GR~\cite{Papanikolaou:2020qtd}.

On the other hand, by fixing $m_\mathrm{PBH}$ and $\Omega_\mathrm{PBH,f}$ we saturated the above inequality in order to find an upper bound on the mass scale $M$ given the fact that the GW amplitude is an increasing function of $M$ as it can be seen by the left panel of \Fig{fig:GW_spectrum_at_evap_different_alpha}. These upper bounds can be seen in the color bar axis of \Fig{fig:M_constraints} but they should not be considered as physical ones given the fact that the value of $M$ is very well fixed by CMB observations. As mentioned before already, given the simplicity of the $R^2$ gravity model we use it as a case study in order not to set precise constraints on $M$ but rather to illustrate that one can use the SIGW portal associated to PBH Poisson fluctuations as a novel probe to constrain alternative gravitational theories.

\begin{figure}[h!]
\begin{center}
\includegraphics[width=0.796\textwidth,  clip=true]
{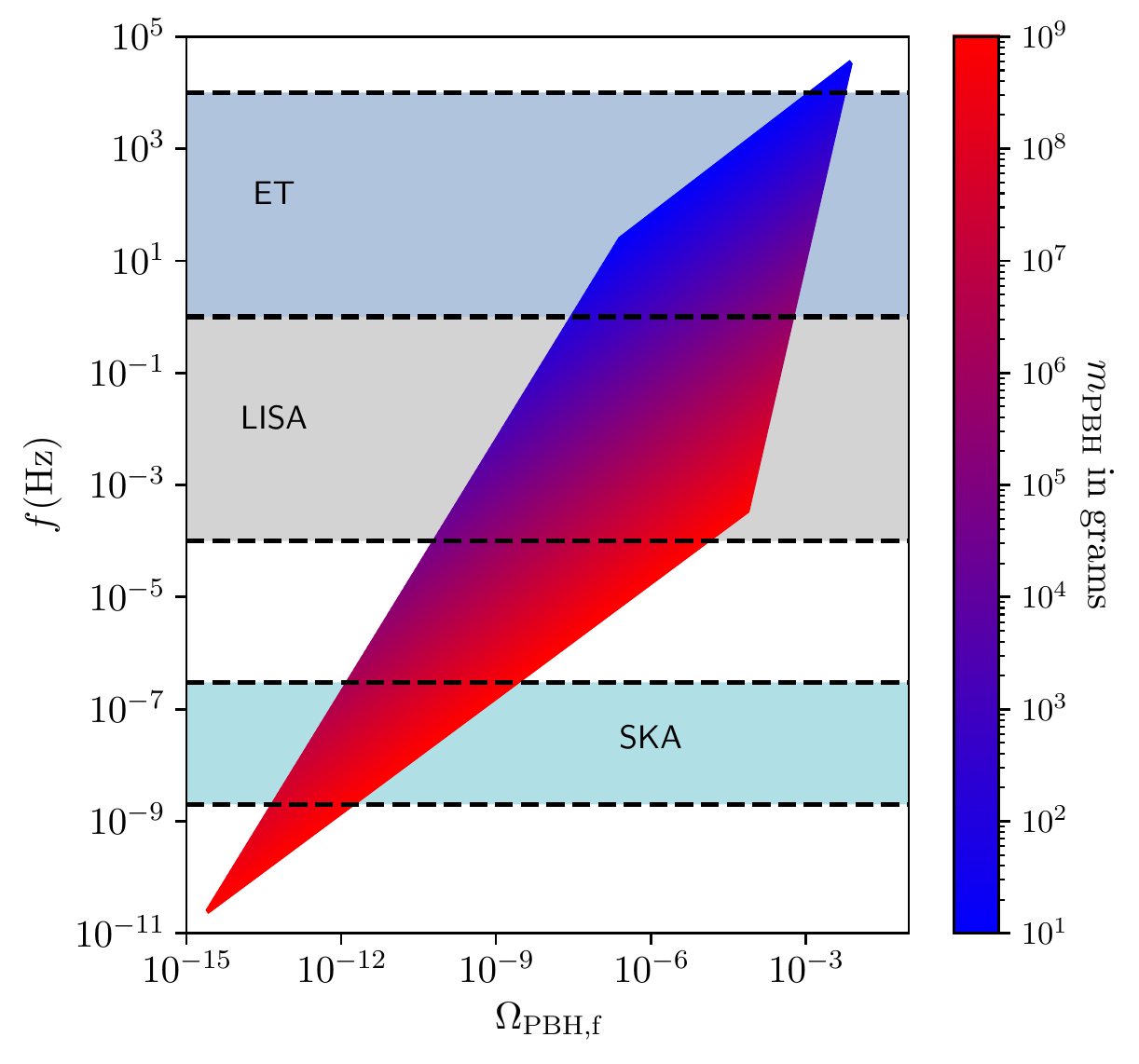}
\caption{The peak frequency of the SIGW signal within $R^2$ gravity produced during an early PBH-dominated era as a function of  
the initial PBH abundance at formation $\Omega_\mathrm{PBH,f}$ (horizontal axis) and the PBH mass $m_\mathrm{PBH}$ (colour coding).
The region of parameter space that is shown corresponds to values of $m_\mathrm{PBH}$ and $\Omega_\mathrm{PBH,f}$ such as that the black holes dominate the energy budget of the Universe for a transient period, see \Eq{eq:domain:OmegaPBHf}, that they form after inflation and Hawking evaporate before big-bang nucleosynthesis, see \Eq{eq:domain:mPBHf}, and that the induced gravitational waves do not lead to a backreaction problem, see \Eq{OmegaPBHf_upper_bound_Starobinsky_conclusions}. For our numerical applications, $g_\ueff=100$, $z_\mathrm{eq}=3387$ and $H_0=70\,\mathrm{km}\,\mathrm{s}^{-1}\,\Mpc^{-1}$. For comparison, the frequency detection bands of ET, LISA and SKA are also shown.}
\label{fig:GW frequency}
\end{center}
\end{figure}

One should comment here on the observational prospects of the aforementioned SIGW signal within the class of $f(R)$ gravity theories. Regarding the frequency of the GW signal these are given by $f\equiv k/(2\pi a_0)$, where $a_0$ is the scale factor today and $k$ is the comoving wavenumber lying within the range $[k_\mathrm{evap},k_\mathrm{UV}]$ with $k_\mathrm{evap}$ being the comoving number crossing the Hubble radius at the PBH evaporation time and $k_\mathrm{UV}$ the UV cut-off scale introduced to avoid entering to the non-linear regime where $\mathcal{P}_\delta(k)>1$. These two comoving wavenumbers depend on the details of the gas of PBHs, namely on their mass $m_\mathrm{PBH}$ and their initial abundance $\Omega_\mathrm{PBH,f}$. Therefore, the only effect of the underlying gravitational theory will be at the level of the upper bound on $\Omega_\mathrm{PBH,f}$ in order to avoid GW overproduction. Consequently, the peak GW frequency at $k_\mathrm{d}$ as a function of $m_\mathrm{PBH}$ and $\Omega_\mathrm{PBH,f}$ can be recast after a straightforward calculation as~\cite{Papanikolaou:2020qtd}
\bea
\label{GW frequency}
\frac{f}{\mathrm{Hz}} \simeq \frac{1}{\left(1+z_\mathrm{eq}\right)^{1/4}}\left(\frac{H_0}{70\mathrm{kms^{-1}Mpc^{-1}}}\right)^{1/2}\left(\frac{g_\ueff}{100}\right)^{1/6}\Omega^{2/3}_\mathrm{PBH,f} \left(\frac{ m_\mathrm{PBH} }{10^9\mathrm{g}}\right)^{-5/6} ,
\eea
where $H_0$ is the value of the Hubble parameter today and $z_\mathrm{eq}$ is the redshift at matter-radiation equality. We show in \Fig{fig:GW frequency} how the SIGW frequency varies with $m_\mathrm{PBH}$ and $\Omega_\mathrm{PBH,f}$. Interestingly, depending on the choice of the PBH mass and the initial PBH abundance, the SIGW frequency can lie within the frequency detection bands of the Einstein Telescope (ET)~\cite{Maggiore:2019uih}, the Laser Interferometer Space Antenna (LISA)~\cite{Audley:2017drz} and the Square Kilometre Array (SKA) facility~\cite{Janssen:2014dka} pointing out the ability of these GW experiments to potentially detect such a signal and measure deviations from GR.

At this point, we should stress that the contribution to the GW spectrum coming from the transition from the PBH dominated era to the radiation dominated one, which in the case of GR and in the regimes of a monochromatic PBH mass function enhances considerably the GW signal as pointed out in~\cite{Inomata:2019ivs,Domenech:2021wkk}, was not considered in this work. This aspect should be considered in future works in order to check which values of GW amplitudes have the potential to be observed by GW experiments.

We close this work by making a comment on the aforementioned procedure. By making use of the cosmological perturbation theory  we extracted the power spectrum of the gravitational potential, and by imposing the UV cut-off scale we ensured that we are well within the perturbative regime. This is very important since it is $\Phi$ that induces the second-order gravitational waves. Nevertheless, from the point of view of the energy density perturbation $\delta$, as it is well established in the context of GR, during matter domination $\delta$ grows linearly with the scale factor. A similar picture was found here too, namely $\delta$ grows with the scale factor although non linearly. Thus, there will be scales where $\delta$ can acquire values larger than one, entering into the non-linear regime although $\Phi$ remains much smaller than one. Therefore, in order to clarify the status of these scales one should follow the full virialisation dynamics \cite{Papanikolaou:2020qtd,Kozaczuk:2021wcl} something which is beyond the scope of this work. However, we may speculate that a growth of $\delta$ will enhance the power spectrum above the Poissonian value, which in turn will lead to an even larger signal than that extracted above. In that sense, the bounds obtained here in particular regarding the initial abundances of PBHs, [see \Eq{OmegaPBHf_upper_bound_Starobinsky_conclusions}] can be considered as conservative ones.

In summary, through the above analysis we showed that the condition to avoid an overproduction of scalar induced gravitational waves associated to PBH Poisson fluctuations at PBH evaporation time can act as a novel method to extract constraints on PBH parameters as well as on gravitational theories, independent from other methods such as the BBN or cosmological confrontations. Hence, by the combined application of all these approaches we can have an improved tool to constrain proposed scenarios and test possible deviations from general relativity.

\begin{acknowledgments}
T.P. acknowledges financial support from the Foundation for Education
and European Culture in Greece and would like to thank as well the Laboratoire Astroparticule and Cosmologie, CNRS Université Paris Cité for giving him access to the computational cluster DANTE where part of the numerical computations of this paper was performed. The authors acknowledge as well the contribution of the COST Action
CA18108 ``Quantum Gravity Phenomenology in the multi-messenger approach''.

\end{acknowledgments}

\appendix

\section{Scalar perturbation equations in $f(R)$ gravity}
\label{AppendixA}
 
In the case of $f(R)$ gravity, in the Newtonian gauge  one extracts the following scalar perturbed field equations \cite{DeFelice:2010aj}:
 \begin{align}
      3\mathcal{H}(\Phi' + \mathcal{H}\Psi) +k^2\Phi &= -4\pi G a^2 \, \delta \rho_\mathrm{tot} \label{P1}, \\
        \Phi' + \mathcal{H}\Psi &= 4\pi G a^2(\bar{\rho}_\mathrm{tot} + \bar{p}_\mathrm{tot}) v_\mathrm{tot},\label{P2} \\ 
         \Phi'' + \mathcal{H}(\Phi' + 2\Psi') + (\mathcal{H}^2 + 2\mathcal{H}')\Phi -k^2(\Phi - \Psi) /3   &= -4\pi G a^2  \,\delta p_\mathrm{tot}, \label{P3} \\
    \Phi - \Psi  &= \, 8\pi G a^2 \, \bar{p}_\mathrm{tot}\Pi_\mathrm{tot} ,\label{eq:Phi-Psi}
    \end{align}
 where 
    \begin{eqnarray}
(\bar{\rho}_\mathrm{tot} + \bar{p}_\mathrm{tot}) \upsilon_\mathrm{tot} \equiv \sum_{l= m,r, f(R)} (\bar{\rho}^{l} + \bar{p}^{l}) v^{l} ,
    \end{eqnarray} 
 and 
       \begin{eqnarray}\label{eq:Pi_tot}
 \bar{p}_\mathrm{tot} \Pi_\mathrm{tot} \equiv \sum_{l= m,r, f(R)} \bar{p}^{l} \Pi^{l}.
    \end{eqnarray} 
    Additionally, the perturbed energy density and pressure of 
    the effective fluid arising from $f(R)$ mortification, are written respectively 
as  
        \begin{align}
 &
 \!\!\!\!\!\!
 \delta \rho_\mathrm{f(R)} \equiv -\delta T^{\mathrm{f(R)} \, 0 }_{0} = -\frac{1}{8\pi G a^2} \Big\{ (1 - F) \big[ -6\mathcal{H}' \Psi + k^2 \Psi - 3\mathcal{H} (\Phi' + \Psi ') - 3 \Phi'' \big] \nonumber 
 \\ 
 &\ \ \ \ \ \ \ \ -3 \mathcal{H}' \delta F + a^2 \delta f/2 - k^2 \Psi + 2k^2 \Phi + 6(\mathcal{H}' +\mathcal{H}^2) \Psi + 3\Phi'' +3\mathcal{H}( \Psi' +3 \Phi') \nonumber  \\ 
  &\ \ \ \ \ \ \ \
  + k^2 \del F + 3\mathcal{H} \delta F ' - 3F' (\Phi' + 2\mathcal{H} \Psi) \Big\},
  \label{dreff}
   \end{align}
     \begin{align}
 &  \!  \delta p_\mathrm{f(R)} \equiv  \frac{\delta T^{\mathrm{f(R)} \, i }_{i}}{3} 
 = \frac{1}{8\pi G a^2}  \Big\{  - ( \mathcal{H}' + 2\mathcal{H}^2) \del F + a^2 \delta f/2   + k^2 (2\Phi - \Psi) + 3\mathcal{H} (\Psi' +3\Phi') \nonumber \\
 &\ \ \ \ \ \ \ \ \ \
+3\Phi'' + 6(\mathcal{H}'\! +\! \mathcal{H} ) \Psi  + \delta F'' + 2k^2 \delta F/3 + \mathcal{H} \delta F ' - F' (2\Phi' + 2\mathcal{H} \Psi + \Psi') -3 \Psi F''  \nonumber \\
 &\ \ \ \ \ \ \ \ \ \
+ (1 - F) \big[ - k^2 \Phi -\Phi''- 3\mathcal{H} (5\Phi' + \Psi ') - (2\mathcal{H}' + 4 \mathcal{H}^2) \Psi     - k^2(\Phi - \Psi )/3 \big]  \Big\} \label{dpeff}.
    \end{align}
    Finally, we have
     \begin{align}
 &(\bar{\rho}_\mathrm{f(R)} + \bar{P}_\mathrm{f(R)})v^\mathrm{f(R)}_{,i} \equiv - \delta T^{\mathrm{f(R)}\, 0 }_{i} = \frac{1}{8\pi G} \Big[ 2(1-F) (\Phi' + \mathcal{H}\Psi)_{,i} + \delta F'_{,i} + F' \Psi_{,i} - \mathcal{H} \delta F_{,i}\Big]   ,
 \end{align}
 and
\beq\label{eq:Pi_f_R}
 \Pi^\mathrm{f(R)}_{ij} \bar{P}_\mathrm{f(R)} \equiv \delta T^{\mathrm{f(R)} \, i }_{j} = \frac{1}{8\pi G a^2} [ (1 - F)( \Phi - \Psi )_{,ij} + \delta F_{,ij}] , \ \ \, i \neq j.
 \eeq
 In this context, we can define the (total) comoving curvature perturbation in the usual manner, namely
 \begin{equation}
     \mathcal{R} \equiv - \Phi - \mathcal{H} \upsilon_\mathrm{tot} \label{R}.
 \end{equation}
 
 \section{The anisotropic stress}\label{app:anisotropic_stress}
\begin{figure}[h!]\centering
{\includegraphics[width=.45\linewidth]{
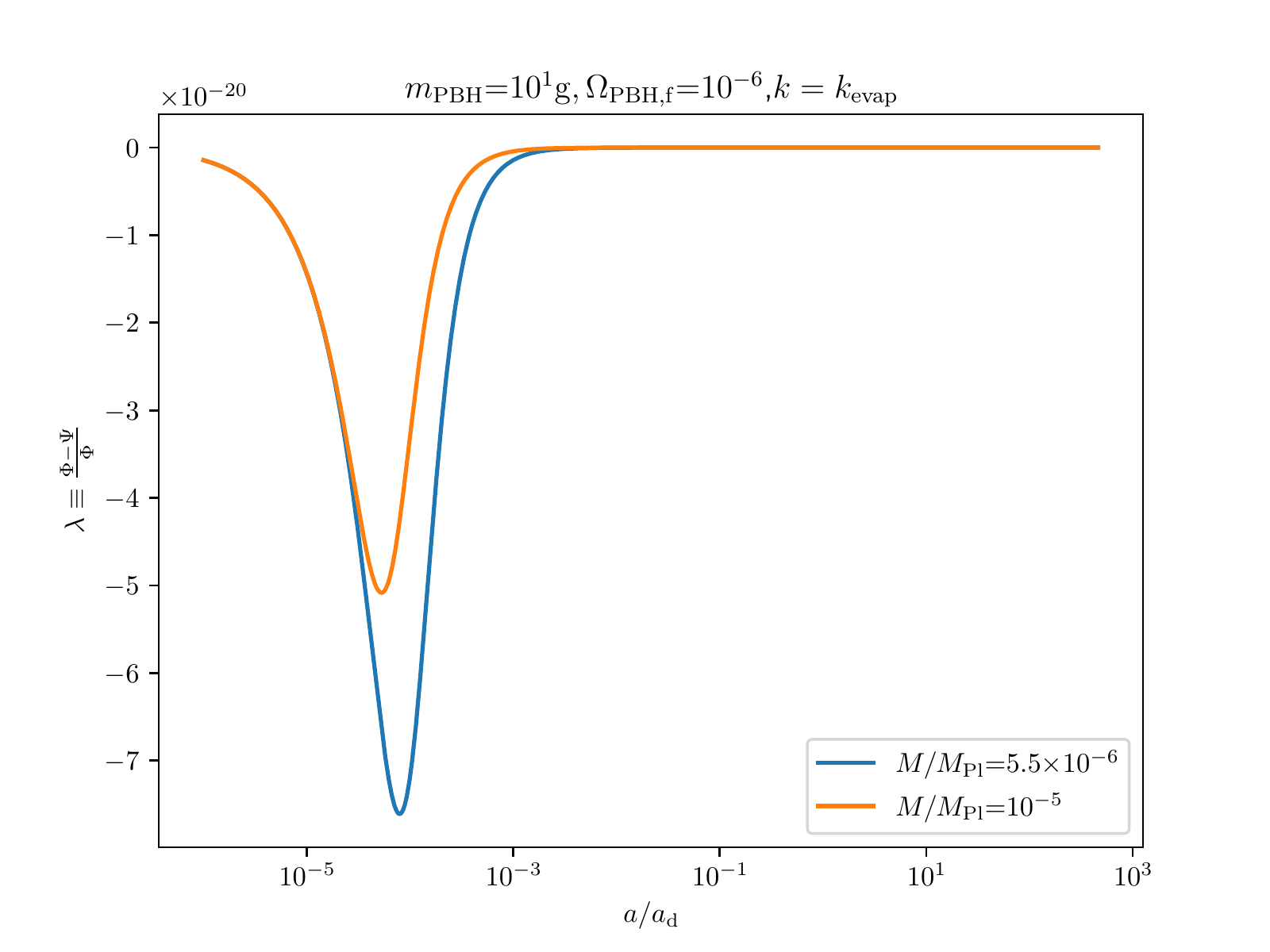}}\hfill
{\includegraphics[width=.45\linewidth]{
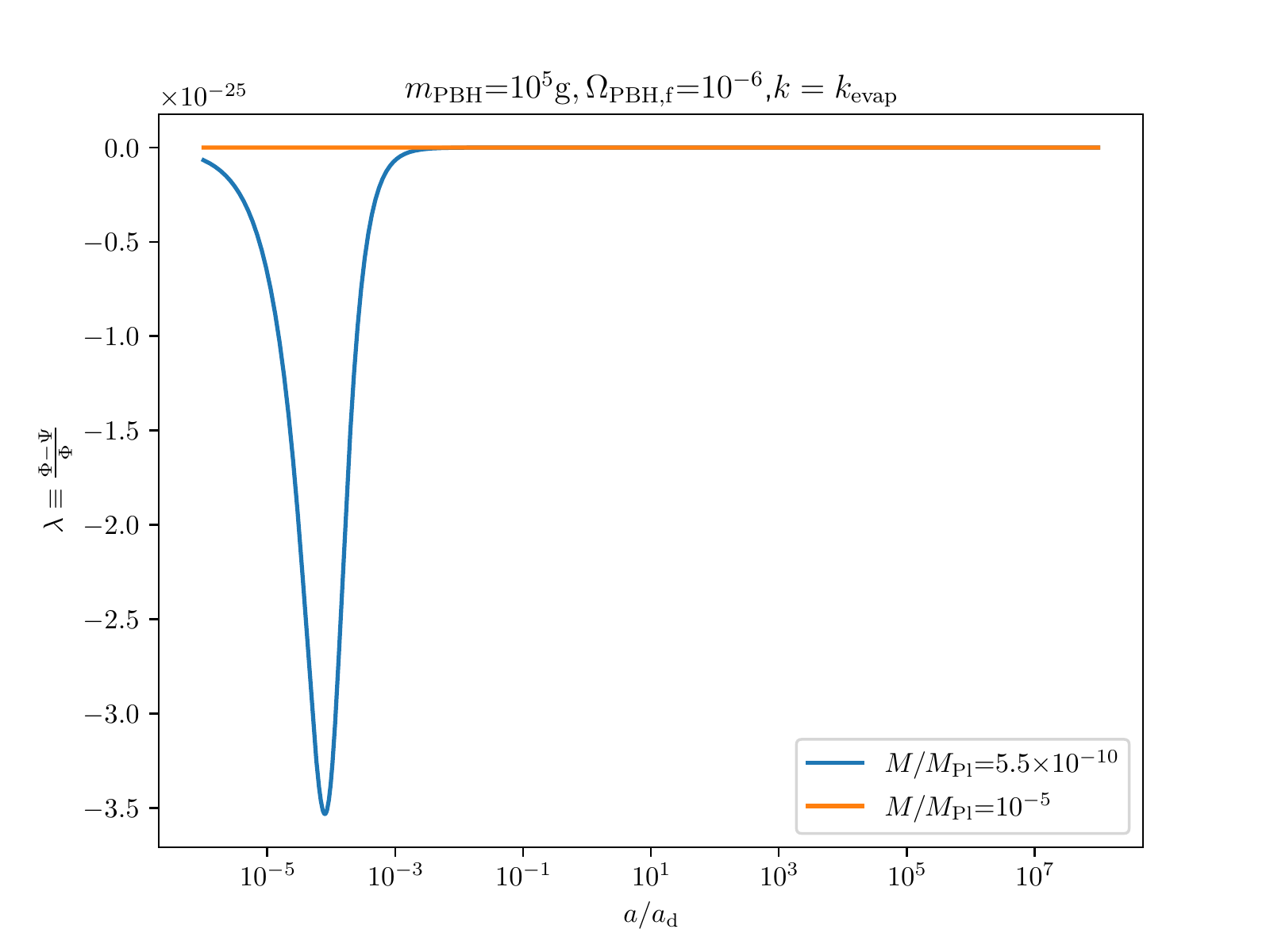}}\par
{\includegraphics[width=.45\linewidth]{
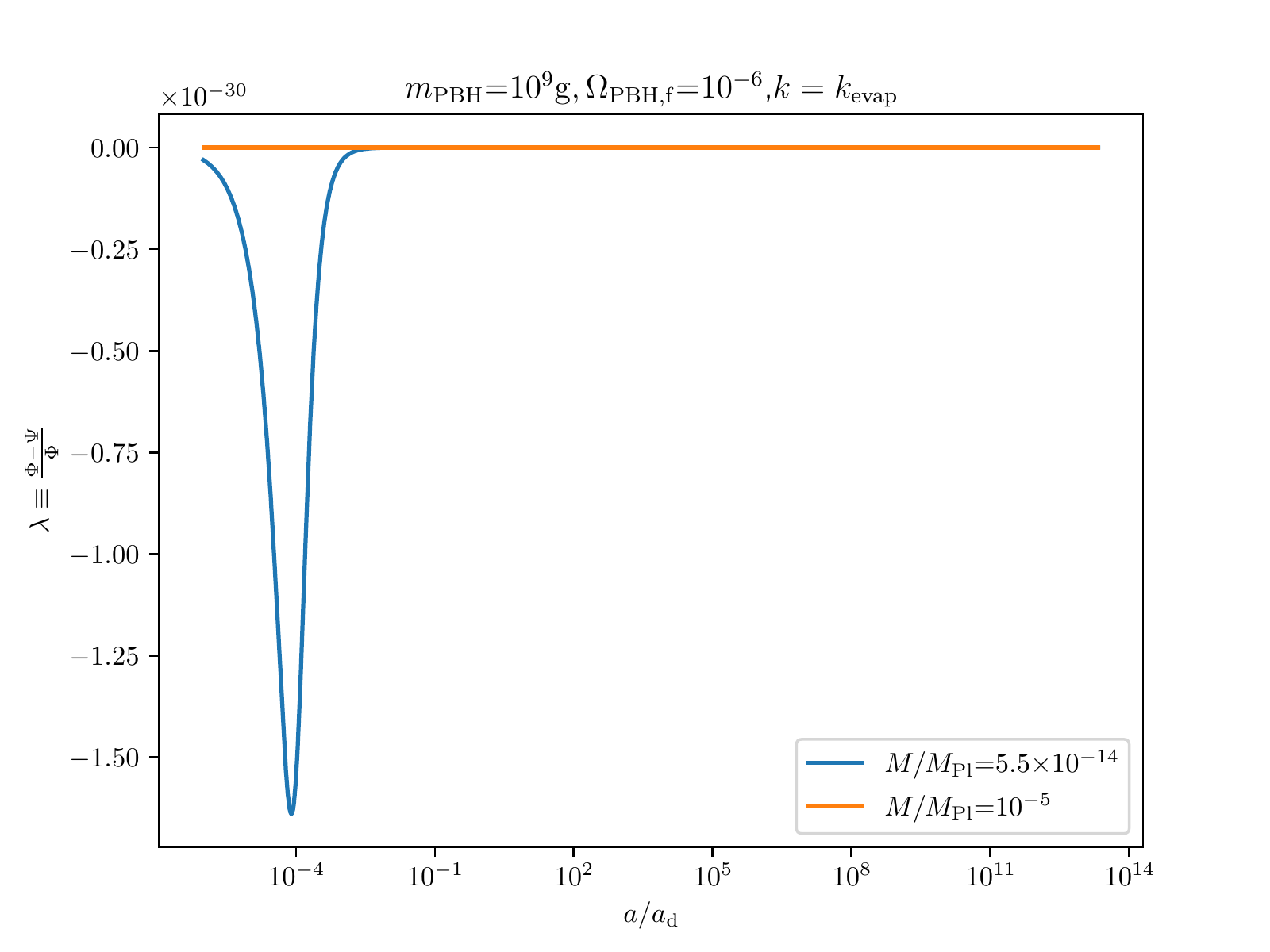}}
\caption{\it{The dimensionless parameter $\lambda\equiv\frac{\Phi-\Psi}{\Phi}$ for $k=k_\mathrm{evap}$ and for different values of the parameter space $(m_\mathrm{PBH},\Omega_\mathrm{PBH,f},M)$. The blue line corresponds to $M=H_\mathrm{f}$ and the orange one to $M=10^{-5}\Mp$. }}
\label{fig:anisotropic_stress_k_evap}
\end{figure}

\begin{figure}[h!]\centering
{\includegraphics[width=.45\linewidth]{
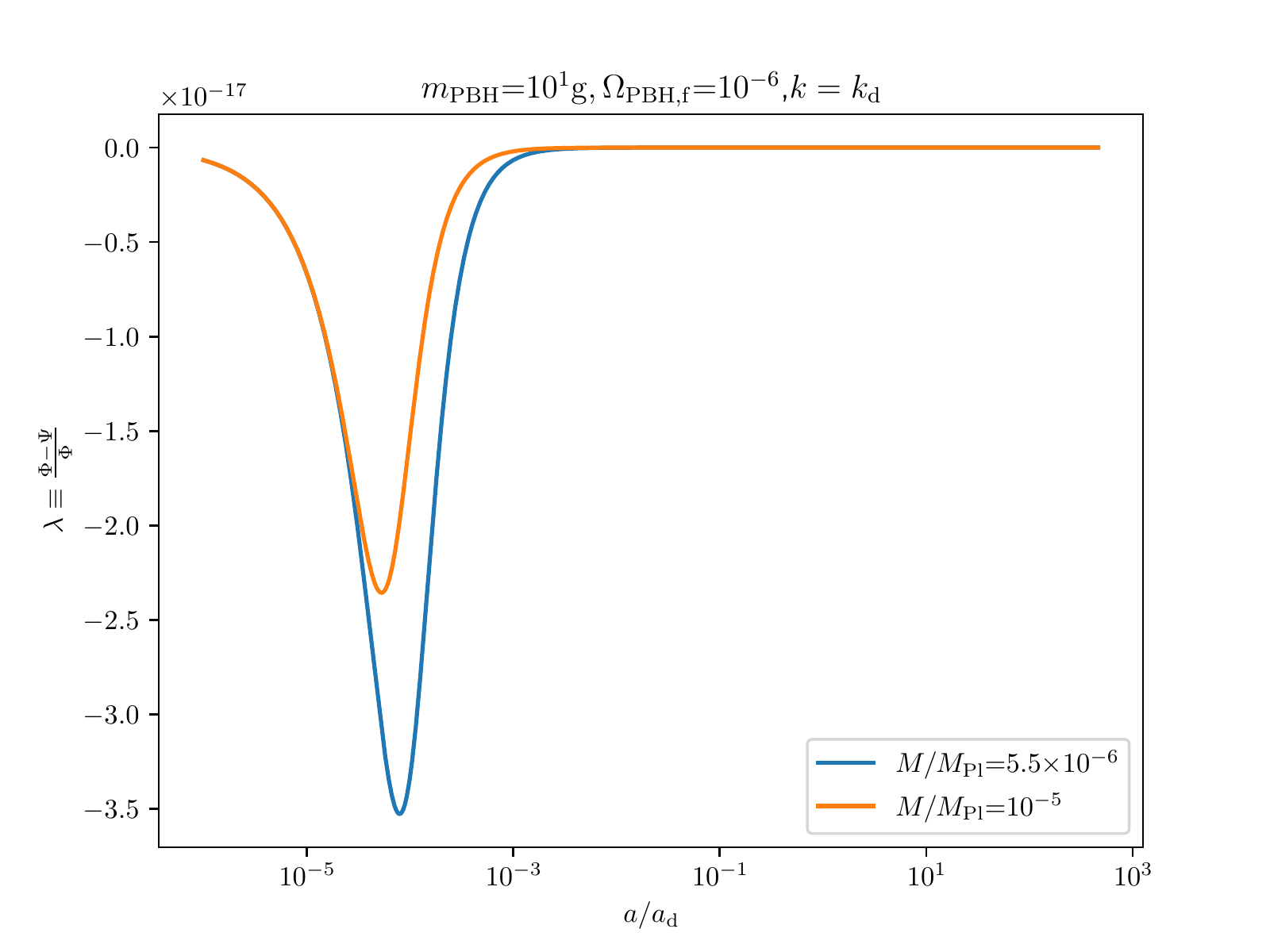}}\hfill
{\includegraphics[width=.45\linewidth]{
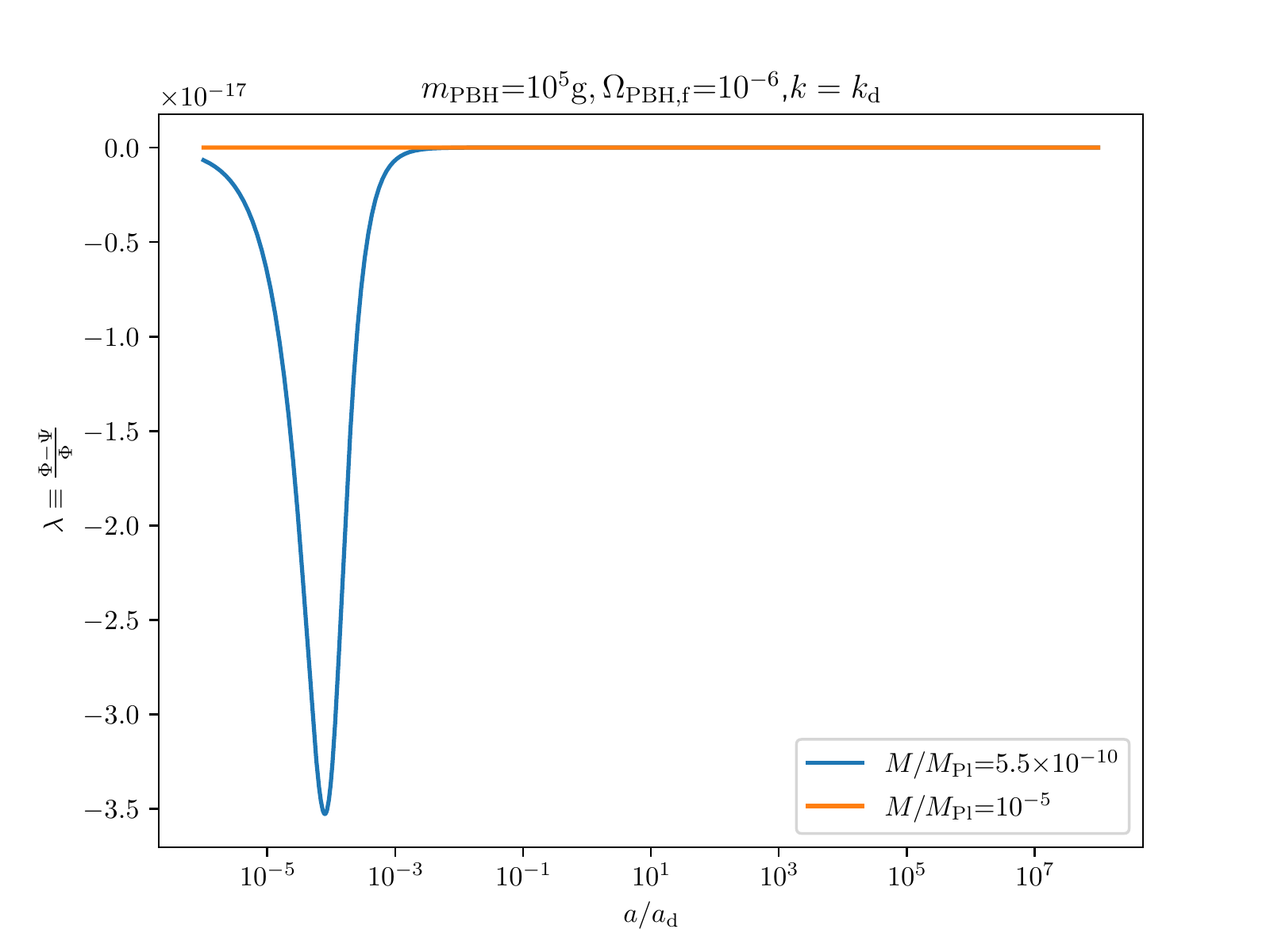}}\par
{\includegraphics[width=.45\linewidth]{
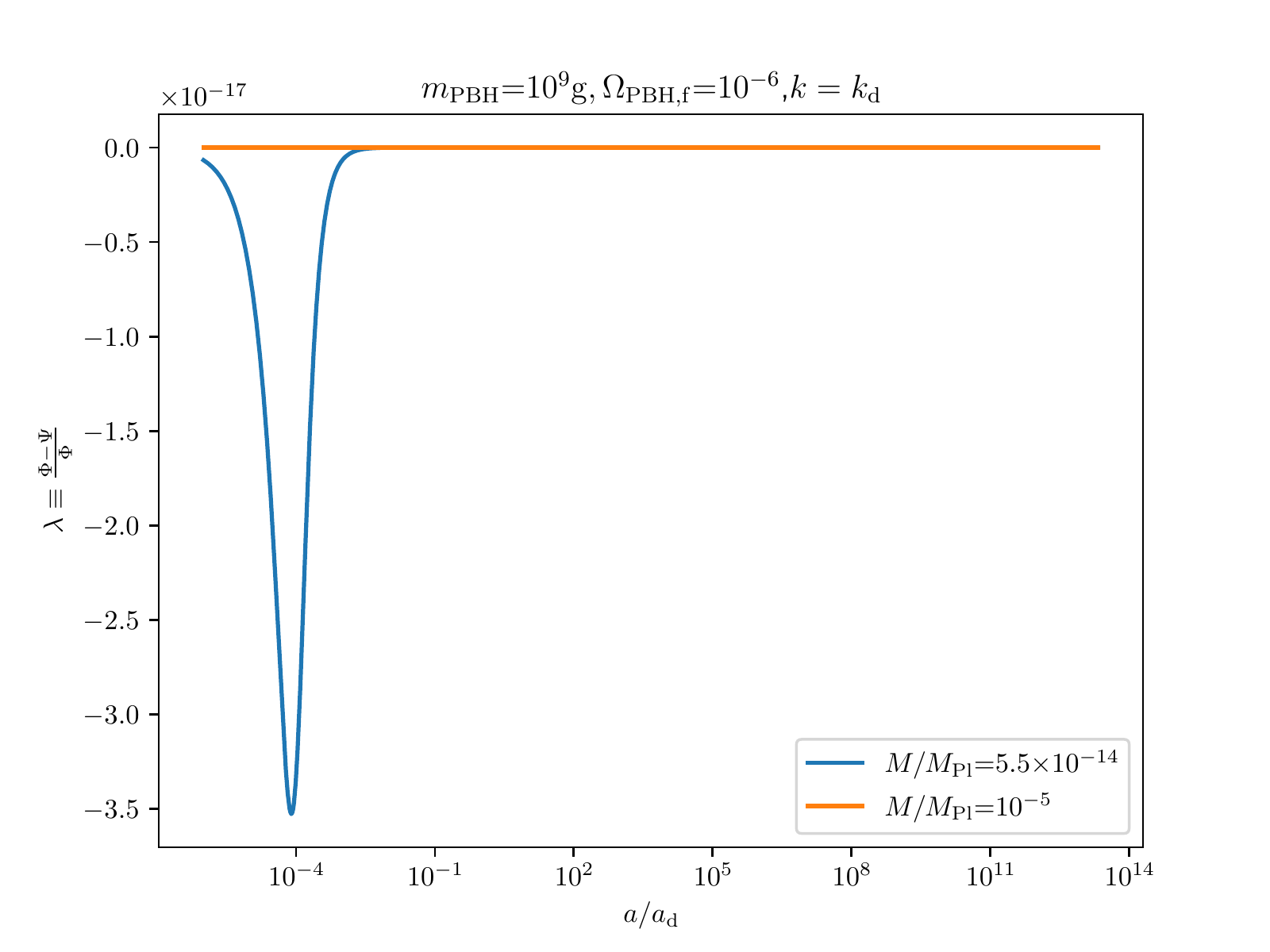}}
\caption{\it{The dimensionless parameter $\lambda\equiv\frac{\Phi-\Psi}{\Phi}$ for $k=k_\mathrm{d}$ and for different values of the parameter space $(m_\mathrm{PBH},\Omega_\mathrm{PBH,f},M)$. The blue line corresponds to $M=H_\mathrm{f}$ and the orange one to $M=10^{-5}\Mp$. }}
\label{fig:anisotropic_stress_k_d}
\end{figure}

\begin{figure}[h!]\centering
{\includegraphics[width=.45\linewidth]{
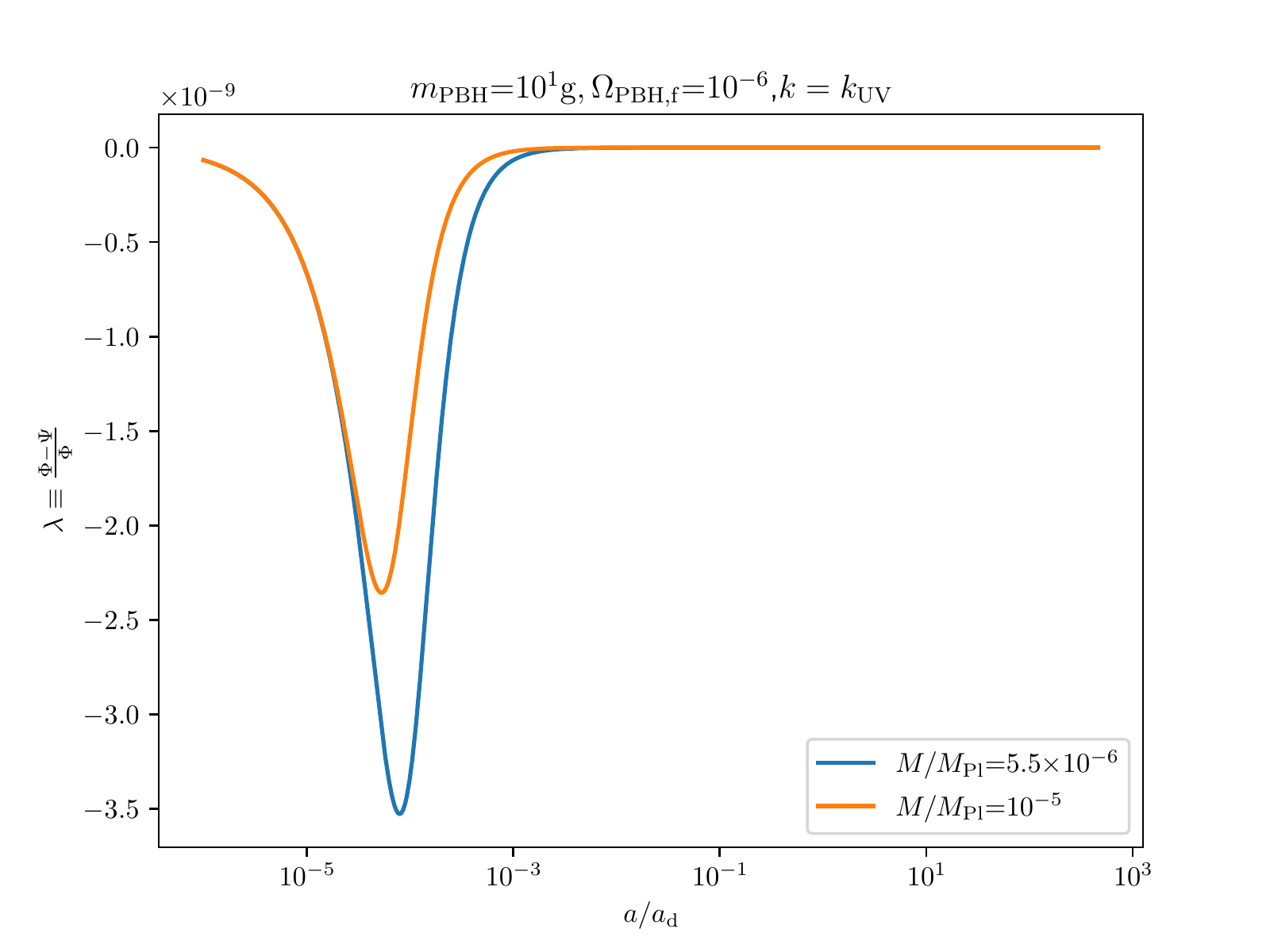}}\hfill
{\includegraphics[width=.45\linewidth]{
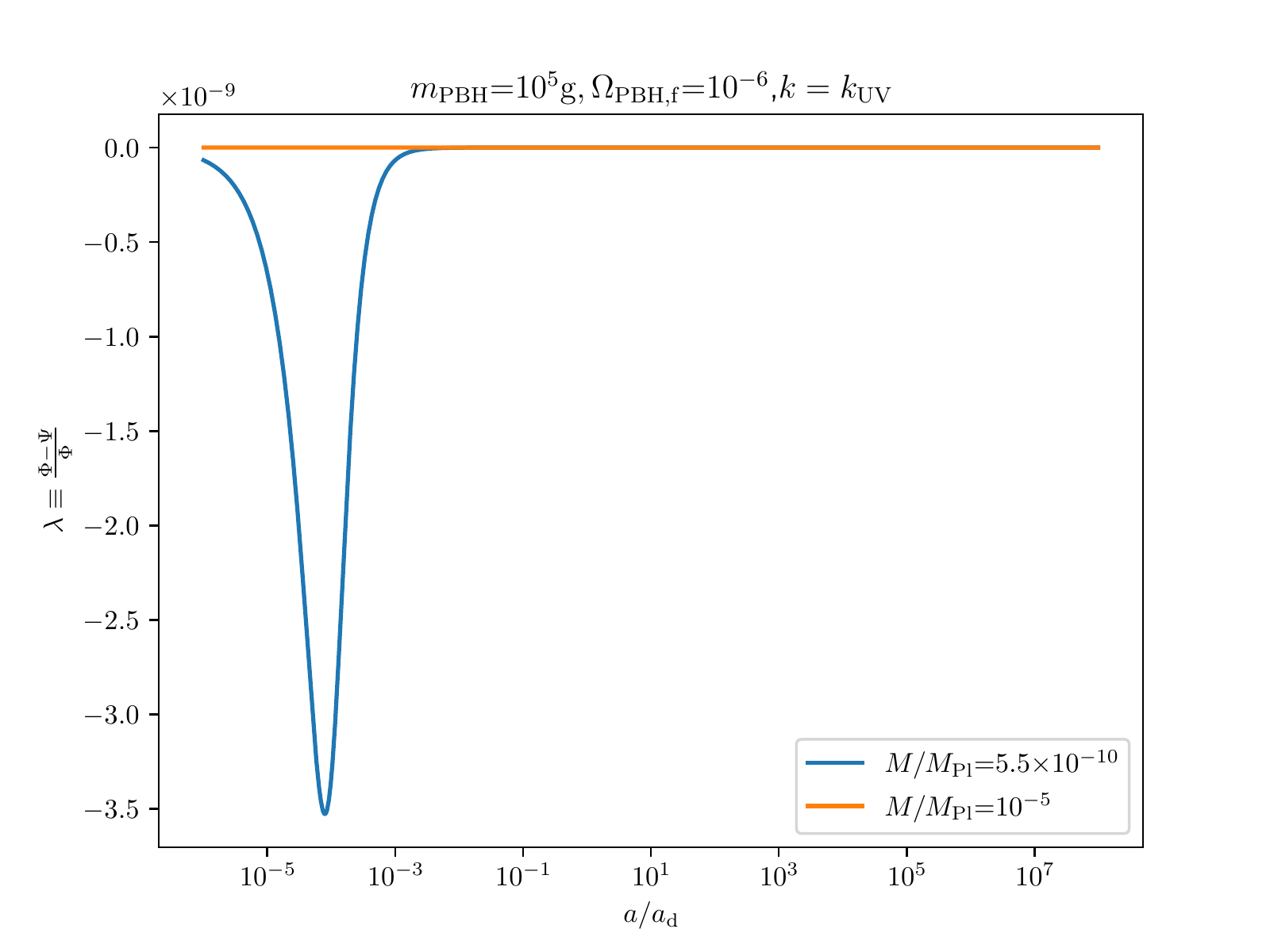}}\par
{\includegraphics[width=.45\linewidth]{
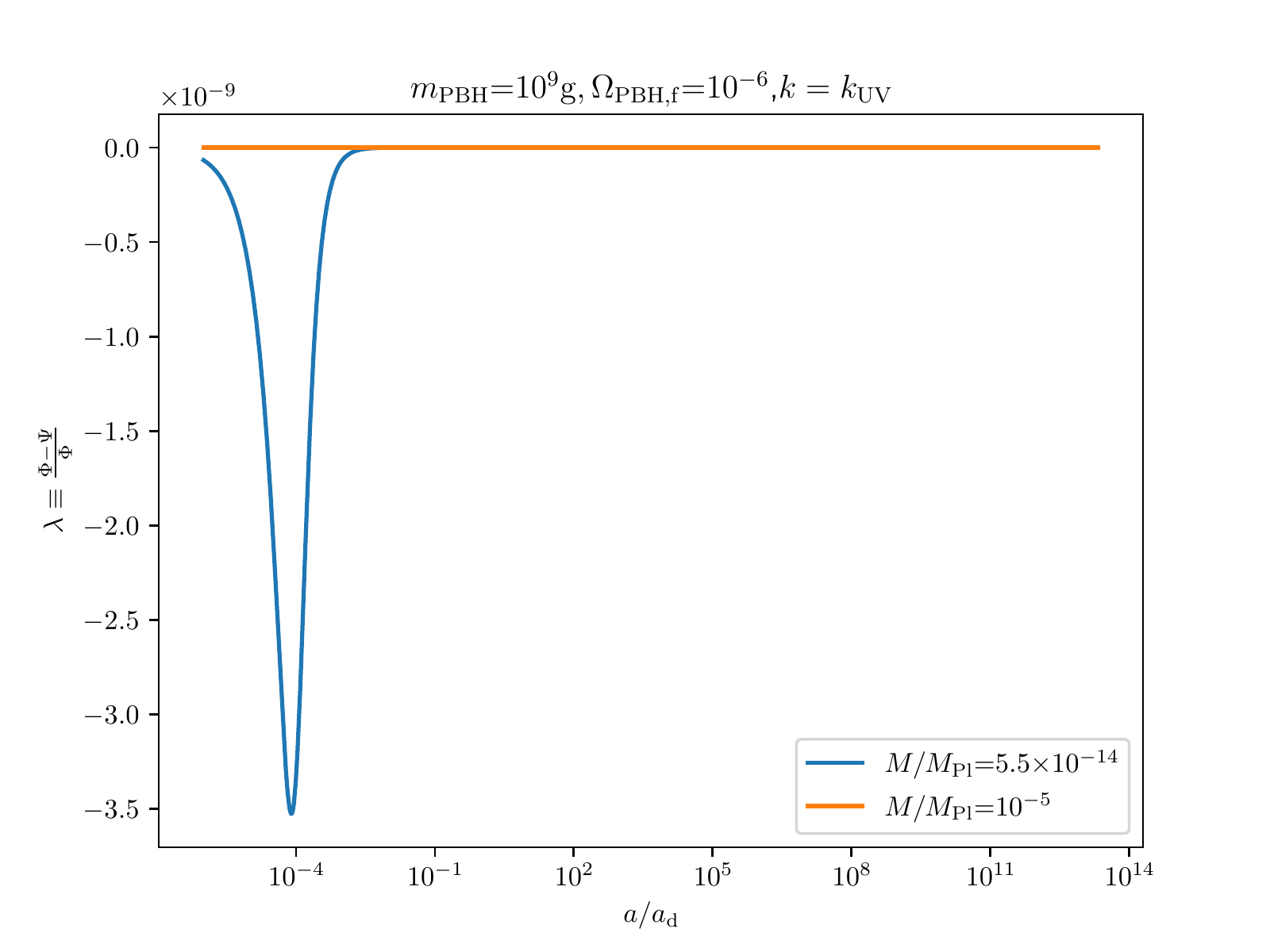}}
\caption{\it{The dimensionless parameter $\lambda\equiv\frac{\Phi-\Psi}{\Phi}$ for $k=k_\mathrm{UV}$ and for different values of the parameter space $(m_\mathrm{PBH},\Omega_\mathrm{PBH,f},M)$. The blue line corresponds to $M=H_\mathrm{f}$ and the orange one to $M=10^{-5}\Mp$. }}
\label{fig:anisotropic_stress_k_UV}
\end{figure}

 Before BBN, which is the period we are interested in, there are no free streaming particles, namely neutrinos or photons, and the dominant matter species is in form of PBHs. Thus, one can safely assume that $\Pi_\mathrm{r}=\Pi_\mathrm{m}=0$. One then is left with the anisotropic stress of the $f(R)$ gravity effective fluid which has a pure geometrical origin. Combining \Eq{eq:Phi-Psi}, \Eq{eq:Pi_tot} and \Eq{eq:Pi_f_R} with $\Pi_\mathrm{r}=\Pi_\mathrm{m}=0$ one can show that 
 \beq\label{eq:Phi-Psi_geometrical}
 \Phi - \Psi = \frac{\delta F}{F},
 \eeq
 with $\delta F = F_{,R}\delta R$ and $\delta R$, being the first order perturbation of the Ricci scalar, given by~\cite{Tsujikawa:2007gd}
 \beq\label{eq:delta_R}
 \delta R = -2 \frac{k^2}{a^2}\frac{\Phi}{1 + 4\frac{k^2}{a^2}\frac{F_{,R}}{F}}
 \eeq
 At this point, one can naturally define a dimensionless quantity denoted here with $\lambda$ as 
 \beq
 \lambda \equiv \frac{\Phi - \Psi}{\Phi},
 \eeq
 which actually quantifies the anisotropic stress of geometrical origin. In the case of $f(R)$ gravity, plugging \Eq{eq:delta_R} into $\delta F = F_{,R}\delta R$  and inserting then $\delta F$ into \Eq{eq:Phi-Psi_geometrical} one can find that 
 \beq\label{eq:gravitional_slip}
 \lambda = \frac{-2\frac{k^2}{a^2}\frac{F_{,R}}{F}}{1 + 4\frac{k^2}{a^2}\frac{F_{,R}}{F}}
 \eeq
 Below we plot this quantity for different values of the wave number $k$ within the range $[k_\mathrm{evap},k_\mathrm{UV}]$, for different values of masses $m_\mathrm{PBH}$ within the range $[10\mathrm{g},10^9\mathrm{g}]$ as well as for different values of the mass parameter of $R^2$ gravity within the range $[H_\mathrm{f},10^{-5}\Mp]$. As one can see from \Fig{fig:anisotropic_stress_k_evap}, \Fig{fig:anisotropic_stress_k_d} and \Fig{fig:anisotropic_stress_k_UV}, $\lambda$ is extremely small signalling that one can safely consider a vanishing  anisotropic stress and as a consequence that $\Phi = \Psi$. Very tiny values of $\lambda$ we also get by varying $\Omega_\mathrm{PBH,f}$
 
 \section{Super-Hubble scales in $f(R)$ gravity} 
 \label{AppendixB}
 
 On super-Hubble scales, \Eq{P1} becomes $ 3\mathcal{H}(\Phi' + \mathcal{H}\Psi) = -4\pi G a^2 \, \delta \rho_\mathrm{tot}$, and thus together with \Eq{P2} yields:
  \begin{equation}
     \mathcal{R} = - \Phi + \frac{\delta_\mathrm{tot}}{3(1+w_\mathrm{tot})} \xrightarrow{(\ref{zeta})} - \zeta, \quad k\ll \mathcal{H}. \label{RZ}
 \end{equation}
Furthermore, from \Eq{R} and \Eq{P2} we can write:
\begin{equation}
    \mathcal{R}  = \Phi + \frac{\mathcal{H}({\Phi}' + \mathcal{H} \Psi)}{4\pi G a^2 \bar{\rho}_\mathrm{tot} (1+w_\mathrm{tot})} \xrightarrow{\mathcal{H}^2 = 8 \pi G a^2 \bar{\rho}_\mathrm{tot}/3}  \Phi + \frac{2}{3} \frac{\Phi'/\mathcal{H} + \Psi}{ 1+w_\mathrm{tot}} \label{Raf}.
\end{equation}
Moreover, from \Eq{eq:Phi-Psi} and \Eq{eq:Pi_tot} we see that $ \Phi - \Psi  =  8\pi G a^2 \, \bar{p}^\mathrm{r}\Pi^\mathrm{r} + \delta F/ F$ and hence by assuming that at super-Hubble modes $\Pi^\mathrm{r} \approx 0$ and $ \delta F \approx 0$ , we deduce that $ \Phi \approx \Psi$. Therefore, under these assumptions and for $k\ll \mathcal{H}$ we can write for $\mathcal{R}$:
\bea
\mathcal{R}  = \frac{2}{3}\frac{{\Phi}^\prime/\mathcal{H}+\Phi}{1+w_\mathrm{tot}}+\Phi\, .
\eea
 
\section{The kernel function $I(u,v,x)$}\label{sec:I(v,u,x)}
In this Appendix we derive the kernel function $I(u,v,x)$ defined in \Eq{I function} for all the three polarization modes,
namely the $(\times)$, the $(+)$ and the scalaron one. In order to achieve this we firstly extract the Green function $G_\boldmathsymbol{k}(\eta,\bar{\eta})$ by solving \Eq{Green function equation}. In particular, \Eq{Green function equation} accepts an analytic solution in the case where $w=0$, which depending on the GW polarization reads as
\beq
kG^{(\times)\;\mathrm{or}\;(+)}_k(\eta,\bar{\eta}) = 
\frac{1}{x\bar{x}}\left[ (1+x\bar{x})\sin(x-\bar{x}) - (x-\bar{x})\cos(x-\bar{x})\right],
\eeq
\beq
\begin{aligned}
kG^{(\mathrm{sc})}_k(\eta,\bar{\eta}) & = \frac{k^3}{x\bar{x}\left(M^2-k^2\right)^{3/2}}\Biggl\{\frac{\sqrt{M^2-k^2}}{k}\left(x-\bar{x}\right)\cosh\left[\frac{\sqrt{M^2-k^2}}{k}\left(x-\bar{x}\right)\right]  \\ & +\frac{(M^2-k^2)x\bar{x}-k^2}{k^2}\sinh\left[\frac{\sqrt{M^2-k^2}}{k}\left(x-\bar{x}\right)\right] \Biggr\}.
\end{aligned}
\eeq
The associated $I(u,v,x)$ function for the $(\times)$ and $(+)$ polarization modes can be recast as
\beq
 \label{I_MD with x_d not zero} 
I^2(x)  = \frac{100}{9}\left[ 1 + \cos(x - x_\mathrm{d}) \left(\frac{3}{x^2}-\frac{3x_\ud}{x^3}-\frac{x_\ud^2}{x^2}\right)
- \sin(x - x_\mathrm{d}) \left(\frac{3}{x^3}+\frac{3 x_\ud}{x^2}-\frac{x_\ud^2}{x^3}\right)
\right]^2 ,
\eeq
and as we can see it does not depend on $u$ and $v$. Similarly,  for the scalaron polarization we have
\beq
 \label{I_MD with x_d not zero-scalaron} 
 \begin{aligned}
I^2(x)  & =\frac{100k^4}{9(M^2-k^2)^6x^6}\Biggl\{
\left(M^2-k^2\right)^2\left[x^3+ M^2x x^2_\mathrm{d} - k^2\left(3x_\mathrm{d} +x(x^2_\mathrm{d}-3)\right)\right] \\ & \times  \cosh\left[\frac{\sqrt{M^2-k^2}}{k}\left(x-x_\mathrm{d}\right)\right]+ k\sqrt{M^2-k^2}\bigl[ M^2x_\mathrm{d}\left(x_\mathrm{d}-3x\right) + \\ & k^2\left(3 +3x x_\mathrm{d}-x^2_\mathrm{d}\right)\bigr]\sinh\left[\frac{\sqrt{M^2-k^2}}{k}\left(x-x_\mathrm{d}\right)\right]
\Biggr\}^2,
 \end{aligned}
\eeq
which is   independent of $u$ and $v$  too.
Taking now into account the fact that $k_\mathrm{UV}=\mathcal{H}_\mathrm{f}\Omega^{1/3}_\mathrm{PBH,f}$ and that roughly  $k<k_\mathrm{UV}$ as well as that $\mathcal{H}_\mathrm{f}\leq M$, then one can easily see that $k/M< \Omega^{1/3}_\mathrm{PBH,f}\ll 1$. Consequently, the above functions in a PBH dominated era and in the subhorizon limit, i.e. $x\gg 1$, become
\beq
I^2(x) = \frac{100}{9}\times
\begin{cases}
1 \mathrm{\;if\;s=(\times),(+)}\\
\frac{k^4}{M^4}  \mathrm{\;if\;s=(\mathrm{sc})}
\end{cases}.
\eeq
 
\bibliographystyle{JHEP} 
\bibliography{PBH}

\end{document}